# Toward Digital Twins in 3D IC Packaging: A Critical Review of Physics, Data, and Hybrid Architectures

Gourab Datta*, Sarah Safura Sharif†, Yaser Mike Banad‡

School of Electrical and Computer Engineering
University of Oklahoma, Norman, 73019, U.S.A.
*gourab.datta-1@ou.edu, †s.sh@ou.edu, ‡ bana@ou.edu,

*Abstract*—Three-dimensional integrated circuit (3D IC) packaging and heterogeneous integration have emerged as central pillars of contemporary semiconductor scaling. Yet, the multiphysics coupling inherent to stacked architectures manifesting as thermal hot spots, warpage-induced stresses, and interconnect aging demands monitoring and control capabilities that surpass traditional offline metrology. Although Digital Twin (DT) technology provides a principled route to real-time reliability management, the existing literature remains fragmented and frequently blurs the distinction between static multiphysics simulation workflows and truly dynamic, closed-loop twins. This critical review distinguishes itself by addressing these deficiencies through three specific contributions. First, we clarify the Digital Twin hierarchy to resolve terminological ambiguity between digital models, shadows, and twins. Second, we synthesize three foundational enabling technologies: (1) physics-based modeling, emphasizing the shift from computationally intensive finite-element analysis (FEA) to real-time surrogate models; (2) data-driven paradigms, highlighting virtual metrology (VM) for inferring latent metrics; and (3) in-situ sensing, the "nervous system" coupling the physical stack to its virtual counterpart. Third, beyond a descriptive survey, we propose a unified hybrid DT architecture that leverages physics-informed machine learning (e.g., PINNs) to reconcile data scarcity with latency constraints. Finally, we outline a standards-aligned roadmap incorporating IEEE 1451 and UCIe protocols to accelerate the transition from passive digital shadows to autonomous, self-optimizing Digital Twins for 3D IC manufacturing and field operation.

## I. INTRODUCTION

Three-dimensional integrated circuit (3D IC) packaging has emerged as a transformative integration strategy, delivering substantial gains in performance, bandwidth, and form factor. In contrast to conventional planar scaling, 3D ICs exploit vertical stacking and short vertical interconnects to reduce communication latency and improve energy efficiency. Industrial deployment is already demonstrating aggressive scaling: for example, TSMC's SoIC® hybrid-bonding technology has reported sub-10 $\mu$m bond pitch for ultra-high-density integration [1]. More recently, the paradigm has expanded toward heterogeneous integration (HI) and chiplet-based architectures, as exemplified by state-of-the-art high-performance computing platforms from AMD [2] and Samsung [3]. While these approaches can improve yield, modularity, and functional density by mixing process nodes and die types, they also introduce unprecedented interconnect complexity spanning dense die-to-die bonds, redistribution layers, and vertical vias thereby intensifying challenges in signal integrity, power delivery, thermo-mechanical reliability, and lifetime management. This shift is accelerating rapidly; the global market for 2.5D and 3D packaging is projected to grow from 5 billion to over 20 billion by 2033 [4]. However, while these architectures maximize functional density, they introduce unprecedented interconnect complexity that threatens to throttle this growth.

The adoption of 3D IC packaging presents a "double-edged sword": while vertical stacking enables ultra-high integration density and shortens critical interconnect lengths, it also introduces tightly coupled failure modes that complicate both design closure and long-term reliability. For example, defect density in advanced packaging has been reported to increase by more than 35% relative to conventional 2D integration, driven in large part by the expanded process flow and the stringent alignment and surface-quality requirements of hybrid bonding and fine-pitch assembly [5]. In parallel, thermal power densities in stacked designs are projected to surpass 1 kW/cm², intensifying temperature gradients and accelerating thermally activated degradation mechanisms, including TSV extrusion and micro-bump cracking [6], [7]. These risks further compound as scaling progresses toward finer micro-bump and TSV geometries: thermo-mechanical mismatch and warpage can trigger delamination and cracking; interconnect fatigue accumulates under power- and temperature-cycling; and electromigration accelerates under elevated current density and temperature. Yet reliability management remains largely dependent on offline physical metrology (e.g., scanning acoustic microscopy (SAM) and X-ray computed tomography (CT)). As noted by Lau (2014) [6], such methods are often throughput-limited, may be destructive or require extensive preparation, and most critically can miss transient or load-dependent defects that manifest only under realistic dynamic thermal and mechanical conditions. The result is a persistent manufacturing "blind spot," in which latent defects in complex 3D stacks escape detection during production and only emerge

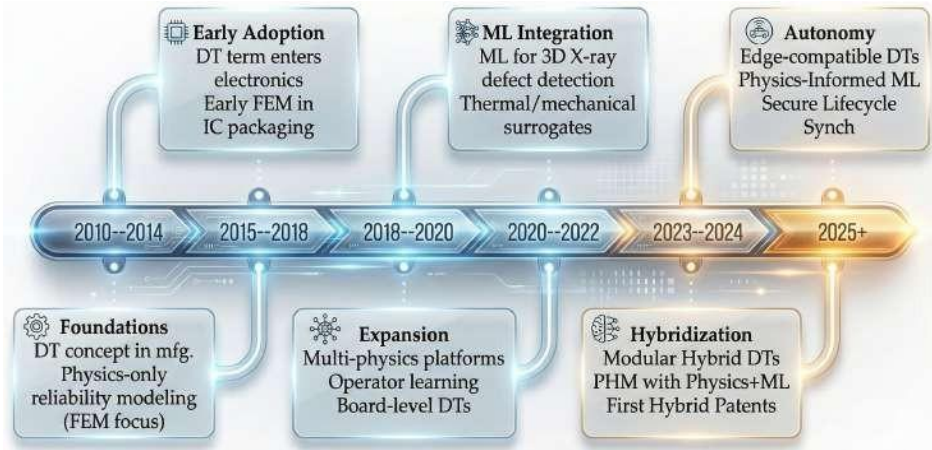

Fig. 1: Historical evolution and projected milestones of Digital Twin in 3D IC packaging, tracing the shift from static physics modeling to autonomous, hybrid frameworks.

after field deployment.

Physics-based modeling has long served as the primary tool for analyzing the tightly coupled thermal, mechanical, and electrical interactions that govern 3D IC behavior. Yet high-fidelity multiphysics approaches (e.g., finite-element and coupled field solvers) are often computationally intensive, difficult to scale to full stacks and system-level contexts, and insufficiently agile when architectures, materials, or process flows change. As emphasized by SEMA (2024) [8], conventional single-physics design methodologies frequently fail to capture the cross-domain couplings that dominate advanced 3D IC operation. Inamdar *et al.* [9] further highlight the potential of Digital Twins (DTs) to connect multiphysics modeling with system-level reliability assessment, underscoring their relevance for next-generation 3D IC packaging. At the same time, non-destructive characterization techniques including 3D X-ray imaging and scanning acoustic microscopy (SAM) increasingly face observability constraints as stacks become denser and interfaces more complex. In ultra-fine-pitch assemblies, limited penetration, resolution, and inspection throughput can restrict in-line detection of hidden defects and weak interfaces [10], [11]. This tension where high-fidelity simulation is too slow for real-time monitoring and control, while physical metrology is too constrained for comprehensive in-line observability creates a widening reliability gap. Bridging this gap motivates a transition from disjoint modeling and inspection pipelines toward a unified Digital Twin framework capable of closed-loop inference, prediction, and decision support.

The Digital Twin (DT) paradigm targets this reliability gap by enabling a bi-directional, continuously updated linkage between the physical package and its virtual representation. As formalized by Shao *et al.* [12], a true DT is distinguished from simpler digital representations by an automated closed-loop control cycle in which in-situ measurements update the virtual model and the model, in turn, informs and optimizes the physical process or operating conditions. A critical examination of the 3D IC packaging literature, however, reveals a frag-mented and often inconsistent use of DT terminology: many reported "digital twins" in advanced packaging operate primarily as *Digital Models* (offline/static simulations) or *Digital Shadows* (one-way monitoring without feedback), rather than as fully closed-loop DTs. Consequently, truly autonomous, closed-loop twins for 3D ICs remain scarce, constrained by three fundamental barriers: (i) a latency bottleneck, wherein high-fidelity multiphysics solvers provide accuracy but are too slow for real-time monitoring and control; (ii) a data sparsity paradox, as data-driven approaches (especially deep learning) thrive on large labeled datasets while defect and failure data in 3D stacks are rare, heterogeneous, and costly to acquire; and (iii) an observability limit, where the lack of standardized, interoperable interfaces for in-situ sensing and telemetry restricts robust acquisition, synchronization, and fusion of measurements across the stack. Fig. 2. summarizes this hierarchy.

The remainder of this article is organized as follows: Section II details the research methodology, outlining the multi-modal search strategy and corpus selection criteria; Section III establishes the theoretical background by defining the Digital Twin hierarchy and reviewing relevant communication standards; Section IV introduces the *Sensing Foundation*, characterizing the observability constraints and in-situ monitoring architectures; Section V examines the *Physics Pillar*, detailing the transition from finite-element analysis (FEA) to machine-learning surrogate modeling; Section VI analyzes the *Data Pillar*, with emphasis on virtual metrology and inferential sensing; Section VII synthesizes these components into a converged *Hybrid Pillar*, proposing a unified architecture grounded in physics-informed neural networks (PINNs); Section VIII presents a comparative analysis and strategic research roadmap; and Section IX concludes the study.

To overcome the limitations of prior surveys and provide a consolidated roadmap for 3D-IC Digital Twins, this article makes four contributions. Firstly Hierarchy clarification. We formalize the distinctions among *digital models*, *digital shad-*

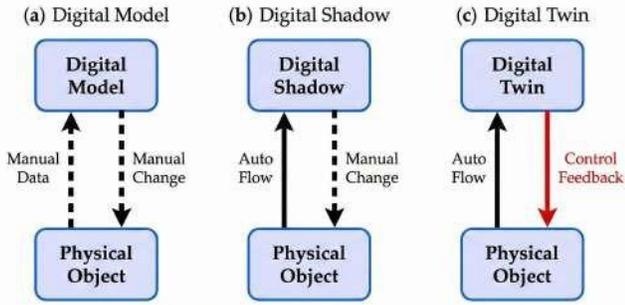

Fig. 2: Data flow hierarchy distinguishing (a) Digital Model (model data exchange), (b) Digital Shadow (automatic monitoring), and (c) Digital Twin (closed-loop automatic feedback/control).

*ows*, and *digital twins* in the specific setting of heterogeneous integration, resolving persistent terminological ambiguity in the packaging and reliability literature. Then **Foundation–Pillars framework.** We synthesize the Digital Twin architectural hierarchy into a unifying framework that identifies a *Sensing Foundation* a multi-modal "nervous system" as the enabling prerequisite for three complementary modeling pillars: (1) the *Physics-Based Pillar* (deterministic white-box modeling), (2) the *Data-Driven Pillar* (stochastic black-box inference), and (3) the *Hybrid Pillar* (converged grey-box architectures). Then **Unified hybrid architecture.** Beyond taxonomy, we propose a converged hybrid architecture that couples Physics-Informed Machine Learning (e.g., PINNs) with modular surrogate models to explicitly address the fidelity latency trade-off required for real-time prediction and decision-making. Finally **Standards-aligned roadmap.** We present a phased research trajectory grounded in industrial standards (e.g., IEEE 1451 and UCIe), charting the progression from passive monitoring to autonomous, lifecycle-aware reliability management.

Because Digital Twin adoption in semiconductor packaging remains at an early stage, the relevant body of work is dispersed across heterogeneous terminology (e.g., "virtual metrology" versus "surrogate modeling") and distributed among multiple research communities. To obtain a comprehensive and reproducible evidence base, we employ a multi-modal search strategy that integrates (i) structured keyword-based retrieval, (ii) AI-assisted semantic discovery with rigorous human verification, and (iii) bibliographic snowballing (as illustrated in Fig. 3).

## II. RESEARCH METHODOLOGY

### A. Search Protocol and Keywords

This work follows a critical review methodology rather than a PRISMA-style systematic review. Primary retrieval was conducted using IEEE Xplore, Google Scholar, and ScienceDirect, spanning the period from 2015 to 2026. To mitigate terminological inconsistency, queries were organized into two logical sets:

- **Set A (Domain Constraints):** ("3D IC" OR "Advanced Packaging" OR "Heterogeneous Integration" OR "Chiplet") AND ("TSV" OR "Hybrid Bonding" OR "Micro-bump")
- **Set B (Methodological Enablers):** ("Digital Twin" OR "Virtual Metrology" OR "Physics-Informed" OR "Surrogate Model" OR "In-situ Sensing")

The initial corpus was constructed by intersecting these sets (i.e., $A \cap B$) to isolate studies at the nexus of advanced packaging and predictive modeling.

### B. Semantic Search and Snowballing

Given that rigid keyword filters can miss relevant contributions that use non-standard labels, we applied two complementary retrieval mechanisms:

1) **AI-Assisted Semantic Discovery and Verification:** Natural-language prompts (e.g., *"machine learning for predicting warpage in stacked dies"*) were used to surface conceptually related studies beyond exact keyword matches. To guard against Generative-AI hallucinations, we enforced a strict *human-in-the-loop* protocol: every AI-suggested title and DOI was manually cross-checked in the publisher's database to confirm existence and bibliographic correctness prior to inclusion.
2) **Bibliographic Snowballing:** We performed backward snowballing from the "Anchor Studies" identified in the initial corpus, screening their reference lists to capture foundational precursors in reliability physics, metrology, and sensor fusion that may predate the explicit "Digital Twin" nomenclature.

### C. Selection Criteria and Corpus Stratification

In contrast to systematic reviews that emphasize statistical completeness, this work follows a *critical review* methodology that prioritizes architectural novelty and transferable design principles over sheer application count. Accordingly, the retrieval and screening process focused on identifying distinct implementation archetypes rather than enumerating incremental variants of conventional offline simulation.

The final bibliography was stratified into two complementary categories:

1) **Core Domain Studies:** Works that explicitly operationalize Digital Twin or Virtual Metrology concepts for 3D IC structures (e.g., TSVs, hybrid bonds). These studies constitute the primary empirical basis for the comparative analysis in later sections.
2) **Theoretical Enablers:** Foundational references that establish conceptual rigor, including general Digital Twin definitions (ISO 23247), machine learning methodologies (e.g., PINNs, Geometric Deep Learning), and sensor physics.

### D. Data Synthesis

The domain-specific corpus was analyzed to extract recurring architectural patterns, which were then mapped to the "Foundation + Pillars" taxonomy. Rather than performing a quantitative meta-analysis, the synthesis focuses on qualitative differentiation, organizing the literature into four distinct domains: the Sensing Foundation (Section IV), the Physics-Based Pillar (Section V), the Data-Driven Pillar (Section VI), and the Hybrid Pillar (Section VII).

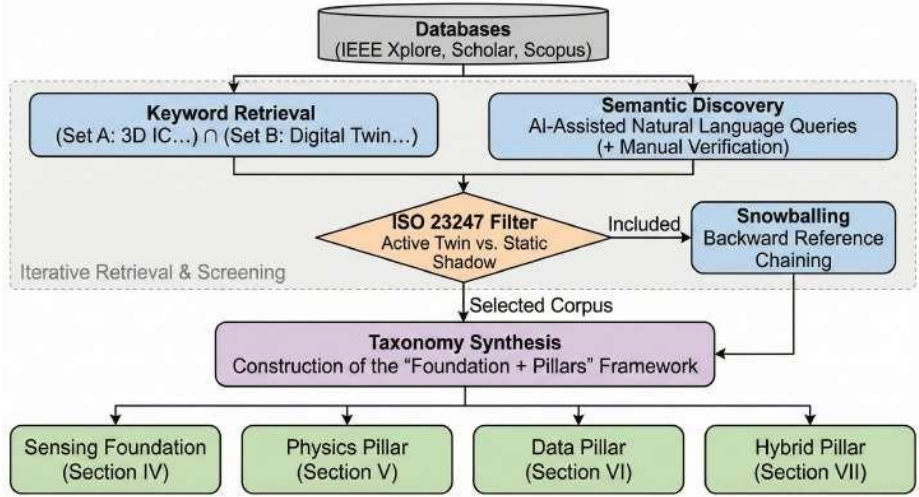

Fig. 3: Schematic representation of the critical review methodology. The workflow integrates standard Boolean keyword retrieval with AI-assisted semantic discovery to mitigate terminology fragmentation in the 3D IC domain. A strict ISO 23247-based filtration step distinguishes active Digital Twins from passive digital shadows, leading to the stratified "Foundation + Pillars" taxonomy synthesized in Sections IV through VII.

### III. BACKGROUND AND CLASSIFICATION OF DIGITAL TWIN ARCHITECTURES

As depicted in Fig. 1, the enabling technologies for 3D IC Digital Twins have evolved rapidly, moving from isolated finite element simulations (2010–2018) to integrated, data-driven frameworks (2018–Present) that pave the way for future self-optimizing systems. The concept of the Digital Twin (DT) was first articulated by Grieves in 2002 in the context of Product Lifecycle Management (PLM) [13] and later formalized by NASA in 2010 for aerospace structural health monitoring [14]; since then, it has evolved into a multidisciplinary cyber physical framework for modeling, simulation, and control of complex engineered systems. The defining characteristic of a true DT in contrast to a conventional offline simulation is its bi-directional data flow: the virtual entity is continuously synchronized with the physical system via real-time sensing, and it returns predictions and control-relevant guidance to optimize operation. To maintain terminological precision in this review, we adopt the ISO-aligned taxonomy proposed by Shao *et al.* [12], which distinguishes three integration levels, namely a Digital Model (static simulation with manual data entry and no automated data flow), a Digital Shadow (one-way, real-time synchronization in which the digital object monitors the physical object), and a Digital Twin (a fully closed-loop system in which the digital representation actively optimizes the physical process). Within 3D IC packaging, DTs are particularly compelling because heterogeneous integration architectures incorporating through-silicon vias (TSVs), hybrid bonding, and chiplets introduce internal interfaces and failure modes that are largely inaccessible to optical inspection. DT technologies mitigate this observability "blind spot" by fusing heterogeneous information streams, including equipment and embedded sensor telemetry (e.g., via IEEE 1451), in-line metrology, and multiphysics models, thereby enabling a shift from reactive failure analysis toward predictive, data-driven quality and reliability management spanning design, assembly, and field operation.

To resolve ambiguity in the literature, we adopt the following strict definitions: Digital Model which is a static, offline simulation (e.g., Ansys/Abaqus) with manual data entry; it supports design verification but lacks operational synchronization. Digital Shadow is a one-way monitoring system where physical state changes are automatically reflected in the digital entity (e.g., dashboard visualization of sensor data), but the digital entity has no direct control over the physical object. Digital Twin is a fully closed-loop system where the digital entity not only mirrors the physical state but also executes autonomous control actions (e.g., process parameter updates) to optimize the physical asset (See Fig. 2). A critical review of recent literature reveals that the majority ($> 80\%$) of reported "Digital Twins" in 3D IC packaging function effectively as Digital Shadows providing advanced monitoring without closing the control loop or as sophisticated Digital Models used for offline reliability assessment; true closed-loop Twins remain an emerging frontier restricted primarily to research prototypes.

*1) Core Architectural Components:* A robust Digital Twin framework for 3D IC packaging comprises three tightly coupled subsystems, as illustrated in Fig. 2:

- **Physical Entity (Asset):** The physical entity denotes the

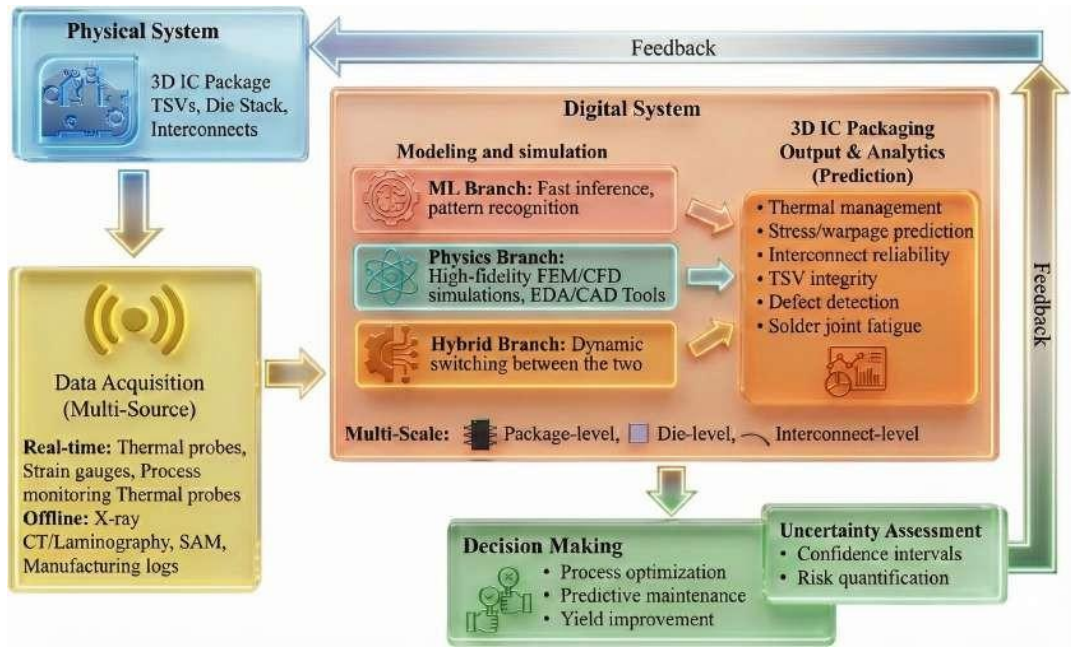

Fig. 4: Operational workflow of the proposed Digital Twin framework for 3D IC packaging. The architecture establishes a closed loop between the Physical System and the Digital System via multi-source data acquisition (real-time sensing and offline metrology). The virtual core integrates three distinct modeling paradigms—Physics-based, Machine Learning (ML), and Hybrid—to generate multi-scale predictive analytics. These insights drive uncertainty-aware decision-making, providing feedback for process optimization and yield improvement.

heterogeneous 3D IC assembly under manufacturing conditions and/or field operation. In contrast to monolithic ICs, it exhibits pronounced multi-scale structure spanning device- and interconnect-level features to package- and system-level components and is governed by strongly coupled thermal, mechanical, and electrical phenomena. Representative physical artifacts include through-silicon vias (TSVs), hybrid-bond interfaces, micro-bumps, redistribution layers (RDL), and thermal interface materials (TIMs), all of which are susceptible to thermally activated aging and thermo-mechanical degradation.

- **Virtual Entity (Model):** The virtual entity is the computational counterpart of the physical asset that reconstructs and predicts its spatio-temporal behavior. Rather than serving as a static repository, it operates as a dynamic modeling and inference engine that may incorporate physics-based formulations (e.g., finite-element or reduced-order models for stress and temperature fields), data-driven components (e.g., statistical learning for defect detection and state estimation), or hybrid surrogate representations that combine mechanistic structure with learned corrections to satisfy real-time latency constraints while preserving physical consistency.

- **Data Services Layer (Interface):** The data services layer provides the bi-directional information pathway that synchronizes the physical and virtual entities. It encompasses signal acquisition, time synchronization, protocol translation, data curation, and multi-source fusion across equipment telemetry, embedded/in-situ sensors, and metrology streams. To ensure interoperability and scalability, this layer should support standardized interfaces for example, IEEE 1451-compliant sensor communication and UCIe-aligned die-to-die connectivity/telemetry so that state estimates and control-relevant directives (e.g., adaptive thermal throttling or dynamic voltage and frequency scaling) can be reliably conveyed back to the physical system.

When these subsystems are co-designed and integrated, the Digital Twin extends beyond passive monitoring to enable closed-loop decision support and control, thereby supporting predictive maintenance, reliability-aware operation, and yield optimization under realistic process and workload dynamics.

The generalized workflow of a digital twin (DT) is depicted in Fig. 4.

### A. Modeling Paradigms for the Virtual Entity

The core predictive and decision-making capability of a Digital Twin is embodied in its *Virtual Entity*. In practice, the choice of modeling strategy is governed by the balance between available physical knowledge, data availability and quality, and the latency constraints imposed by monitoring and control loops. Accordingly, Virtual Entity modeling approaches are commonly categorized into three paradigms physics-driven (white-box), data-driven (black-box), and hybrid (grey-box) as illustrated in Fig. 5.

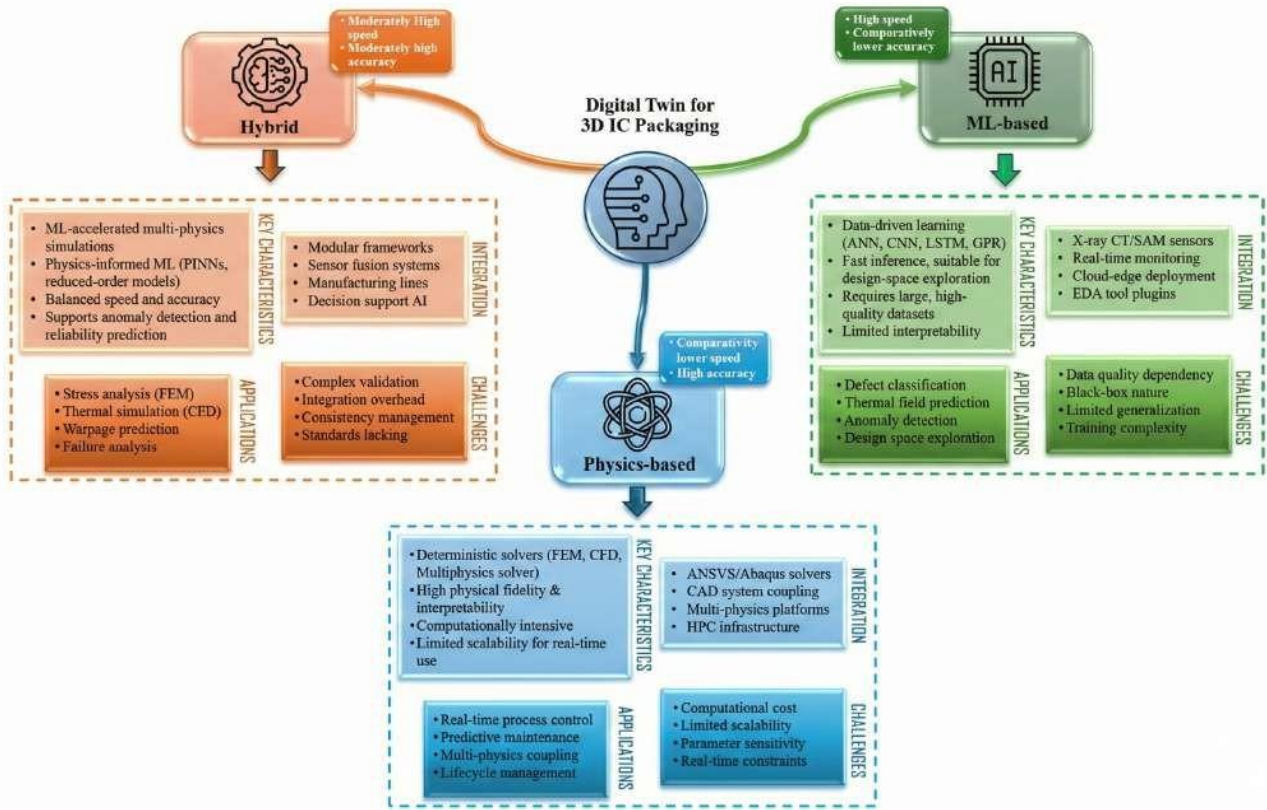

Fig. 5: Classification of modeling paradigms for the Virtual Entity of 3D IC Digital Twins. The framework categorizes architectures into three domains: (a) Physics-based (White Box), characterized by high fidelity and interpretability but high computational cost; (b) ML-based (Black Box), offering high-speed inference for design exploration but limited by data dependency; and (c) Hybrid (Grey Box), which leverages Physics-Informed ML (PINNs) and modular frameworks to balance speed and accuracy. For each paradigm, the diagram details specific key characteristics, integration requirements, target applications, and technical challenges.

*1) Physics-Driven Paradigm (White Box):* Physics-driven approaches employ explicit mathematical formulations most commonly coupled partial differential equations (PDEs) to represent governing mechanisms such as heat transport, thermomechanics, and electromagnetics. In 3D IC packaging, these models are typically evaluated using high-fidelity numerical solvers (e.g., finite-element analysis (FEA) and computational fluid dynamics (CFD)) to resolve temperature fields, stress/strain tensors, warpage, and related multi-physics interactions.

- **Strength:** High interpretability and strong extrapolation potential. Because the model structure is anchored in first principles (e.g., constitutive relations and conservation laws), it can provide physically consistent predictions in operating regimes not explicitly represented in empirical datasets.
- **Limitation:** The latency bottleneck. Fully coupled, high-resolution multi-physics simulations at package or system scale can require hours to days per evaluation, which is incompatible with real-time inference and closed-loop control requirements in manufacturing and field operation [8].

*2) Data-Driven Paradigm (Black Box):* Data-driven approaches learn an input–output mapping (e.g., $y = f_\theta(x)$) directly from observed data without explicitly encoding governing physics. In the context of 3D IC packaging, such models have been applied to tasks ranging from unsupervised anomaly detection in sensor streams (e.g., autoencoders) [15] to supervised regression and classification for estimating stress, warpage, or defect states from process signatures and metrology [16].

- **Strength:** Low-latency inference (often milliseconds) and the capacity to exploit high-dimensional, heterogeneous sensor and process data, including effects that are difficult to parameterize in purely mechanistic models.
- **Limitation:** Lack of physics compliance and limited out-of-distribution generalization. Black-box predictors can violate basic physical constraints (e.g., non-negativity, conservation laws) when training data are sparse, biased, or noisy, and their performance can degrade sharply under process drift or novel architectures outside the training distribution [17].

*3) Hybrid Paradigm (Grey Box):* Hybrid paradigms aim to reconcile the complementary strengths of mechanistic modeling and data-driven learning by coupling physics-based structure with learned components that provide speed, calibration, and adaptability. Two widely adopted mechanisms are:

- **Physics-Informed Machine Learning (PIML):** Incorporating physical laws and constraints (often PDE residuals and boundary/initial conditions) directly into the learning objective or architecture, exemplified by Physics-Informed Neural Networks (PINNs) [18].
- **Residual (Discrepancy) Modeling:** Using a physics model to predict nominal behavior and training a learned model to estimate the residual error (model-form uncertainty) from real-time measurements, thereby correcting systematic bias and unmodeled effects [19].

Hybrid models are particularly well suited to 3D IC Digital Twins because they can preserve physical consistency under data scarcity while meeting stringent latency requirements for in-line monitoring and process control [20].

## IV. THE SENSING FOUNDATION: THE DIGITAL TWIN NERVOUS SYSTEM

Before a Digital Twin can model, predict, or control a 3D IC package, it must first *observe* it. Yet, as illustrated in Fig. 6, the architectural complexity of advanced heterogeneous integration imposes a fundamental *observability limit*: critical internal features including hybrid-bond interfaces, TSV sidewalls, underfill integrity, and localized thermal hotspots are buried within the stack and are therefore inaccessible to conventional surface-level metrology. As a result, the Sensing Foundation is not merely a data conduit; it serves as the Digital Twin's "nervous system," responsible for acquiring, synchronizing, and fusing multi-modal telemetry across extreme spatiotemporal scales from nanosecond-scale on-chip voltage droops to hour-long volumetric X-ray scans to reconstruct the package's latent internal state and provide reliable inputs for the subsequent modeling pillars as illustrated in Tabel I.

### A. The Observability Limit

The fundamental barrier to accurate 3D IC modeling is the *observability limit*. Critical reliability interfaces such as hybrid bond pads and TSV sidewalls are buried deep within the stack, rendering them inaccessible to standard optical inspection. As emphasized by Varshney et al. [21], this creates a "metrology gap" where non-destructive testing (NDT) faces a rigid trade-off: high-resolution techniques like Nano-CT are too slow for inline control, while faster acoustic methods lack the resolution for dense interconnects[23]. The Digital Twin must therefore act as a "soft sensor," fusing sparse high-fidelity NDT data with continuous indirect telemetry to reconstruct the hidden state of the package.

### B. Multi-Modal Sensing Architecture

To close this loop, the sensing architecture operates across three hierarchical levels:

*1) Tool-Level Telemetry (Process Context):* Equipment sensors provide the "boundary conditions" for the Twin. Modern plasma etch and reflow tools stream high-frequency variables (chamber pressure, RF bias, gas flow). While these do not measure the wafer directly, Virtual Metrology (VM) [24] models use them to predict latent defects, such as TSV scallop depth or underfill voiding probability, before physical inspection occurs.

*2) Embedded On-Chip Sensing (Real-Time State):* For operational Digital Twins, static NDT is insufficient. The package must self-report its health using embedded structures:

- **Thermal Profiling:** Distributed thermal sensors (DTS) based on BJT or diode structures are standard for mapping heat flux. However, in logic-on-logic stacks, heat trapped in the middle dies ("dark silicon") often evades surface detection, necessitating 3D-aware sensor placement algorithms[25].
- **Mechanical Stress Monitors:** CMOS-compatible piezoresistive stress sensors [26] (n-type/p-type rosettes) are critical for monitoring package warpage during reflow. These sensors allow the Digital Twin to validate Finite Element models of die cracking in real-time.
- **Emerging Nanomaterials:** As detailed by Banadaki et al. [22], research into graphene nanoribbon (GNR) sensors offers a path to sub-micron thermal and strain sensing within the BEOL layers, potentially bypassing the size constraints of traditional analog sensors.

### C. Data Fusion Challenges

The integration of these diverse streams introduces the challenge of bandwidth mismatch. A Digital Twin must reconcile MHz-rate voltage signals from ring oscillators with quasi-static mechanical creep measurements. This necessitates "edge-heavy" architectures (discussed in Section VI) where high-speed data is compressed locally, transmitting only feature vectors (e.g., peak stress events) to the central Twin to avoid saturating the communication bus.

## V. THE PHYSICS PILLAR: DETERMINISTIC MULTIPHYSICS MODELING

With the *Sensing Foundation* established in Section IV, the critical question becomes: how does the Digital Twin translate these raw observations into predictive insights? This task falls to the Virtual Entity, which relies on three distinct modeling paradigms. We begin with the first of these: the Physics-Based Pillar. Physics-based Digital Twins adopt a *white-box* modeling paradigm in which the virtual entity is anchored in explicit domain knowledge and first-principles formulations, rather than learned solely from empirical correlations. Instead of approximating behavior through statistical mappings, physics-based twins resolve the coupled governing equations of heat transfer, solid mechanics, and electromagnetics (and, when relevant, fluid flow) to estimate and predict the evolving state of the physical system. In 3D IC packaging, where reliability is dominated by strongly coupled interactions among heterogeneous materials and interfaces (e.g., Cu TSVs, Si interposers, low-$k$ dielectrics, polymer underfills, and solder interconnects), such high-fidelity modeling remains indispensable for interrogating failure mechanisms that are difficult or impossible to observe directly with in-line sensing.

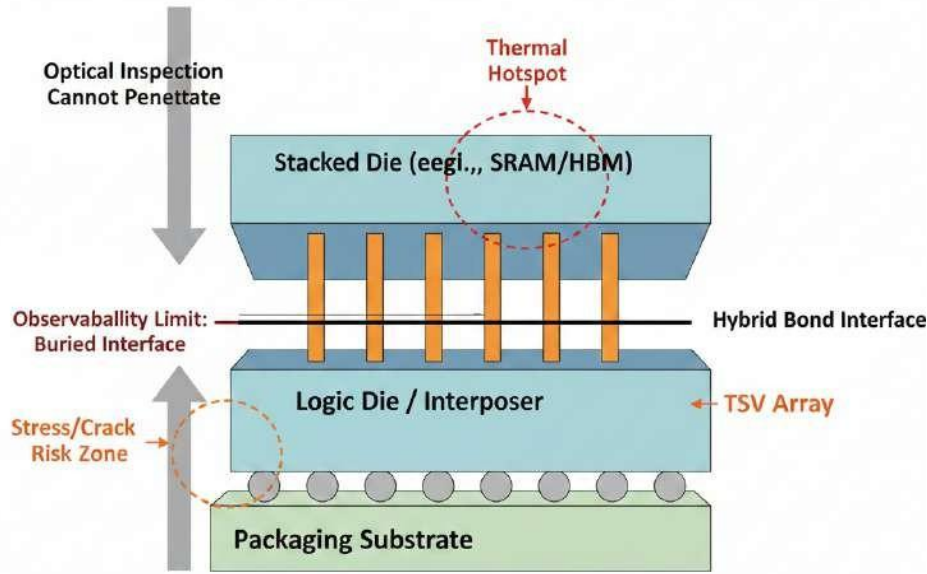

Fig. 6: Schematic cross-section of a representative 3D IC heterogeneous package illustrating key architectural features and associated reliability challenges. The complex, multi-layered stacking of logic and memory dies via Through-Silicon Vias (TSVs) and Hybrid Bonding creates numerous "buried" interfaces. These deeply embedded features result in severe observability limits for physical sensors, creating critical thermal hotspots and stress concentration zones that can only be effectively monitored and managed through a comprehensive Digital Twin framework.

This section surveys recent progress in physics-based Digital Twin frameworks for 3D IC packaging. We first establish architectural requirements for deterministic, high-fidelity simulation in a closed-loop setting and then discuss representative applications in warpage prediction, thermo-mechanical reliability assessment, and fracture/delamination mechanics.

### A. Architecture of the Physics-Based Twin

A physics-based Digital Twin mirrors the physical asset through deterministic, equation-driven simulation; however, unlike a static *Digital Model*, a Twin must support time-evolving synchronization with the physical system. Practically, this requires an explicit *model update mechanism* that adapts boundary conditions, source terms, and effective material/interface parameters in response to measurements and inferred degradation (e.g., increased thermal contact resistance due to interfacial delamination, stiffness reduction due to cracking, or creep-induced constitutive parameter drift).

A robust architecture typically comprises the following subsystems (Fig. 7):

- **Physical Entity:** The 3D IC package (during manufacturing and/or field operation), subject to time-varying workload-induced power dissipation, stochastic usage patterns, and environmental stressors such as ambient temperature fluctuations and board-level constraints.

- **Computational Core:** The high-fidelity multiphysics engine (e.g., FEM/FEA for thermo-mechanics, CFD where convection is relevant, and electromagnetic solvers for current/field distributions). For advanced packaging, the core must accommodate nonlinear and history-dependent constitutive behavior for example, Anand-type viscoplasticity for solder creep, viscoelasticity for polymeric underfills, contact and frictional interface laws, and temperature-dependent material properties to capture fatigue accumulation and stress evolution under realistic thermal and mechanical cycling.

- **Dynamic Input Interface:** A pipeline for ingesting time-varying inputs and health indicators, including spatial power maps (die- and chiplet-level), boundary conditions (e.g., convection coefficients, heat-sink conditions), assembly constraints, and degradation proxies derived from inspection or telemetry (e.g., crack-growth indicators, delamination extent, or interface-property changes).

- **Synchronization and State/Parameter Update Layer:** The assimilation mechanism that reconciles simulation predictions with observed behavior by estimating latent states and calibrating uncertain parameters. For example, if in-situ sensing indicates a sustained temperature mismatch relative to the simulated field, the update layer can adjust thermal contact resistance and/or effective material properties to restore consistency; similarly, warpage measurements can be used to refine package-level mechanical constraints or constitutive parameters.

TABLE I: Taxonomy of sensing modalities for 3D IC Digital Twins: from process tools to on-chip telemetry.

| Sensor Category | Specific Modality | Measurand | Bandwidth/Sampling | Integration Level | Key Challenges for Digital Twins |
|---|---|---|---|---|---|
| Process Telemetry | OES / RF probes | Plasma density, etch rate | 10 Hz–1 kHz | Equipment (chamber) | Indirect correlation to on-wafer features; noisy signals. |
| Structural NDT | 3D X-ray CT | Interconnect voids, cracks | Offline (hours/scan) | Package (ex-situ) | **Observability limit:** trade-off between voxel resolution and throughput [21]. |
| | Scanning acoustic microscopy (SAM) | Delamination, underfill voids | Offline (minutes/scan) | Package (ex-situ) | Blind to high-impedance multilayer stacks (shadowing). |
| On-Chip (FEOL) | Thermal diodes / BJT | Junction temperature ($T_j$) | 1 kHz–1 MHz | Active die (Si) | Limited spatial granularity; cannot sense BEOL/interposer heat directly. |
| | Ring oscillators (RO) | IR drop, NBTI aging | > 10 MHz | Active die (Si) | Distinguishing voltage noise from thermal drift. |
| | Piezoresistive rosettes | Mechanical stress ($\sigma_{xx}$, $\sigma_{yy}$) | 10 Hz–100 Hz | Active die (Si) | Requires large footprint; extensive temperature compensation needed. |
| Emerging (Research) | Graphene nanoribbons | Local strain / hotspots | High (>GHz potential) | BEOL / interposer | Integration compatibility with CMOS process flows [22]. |

- **Verification, Validation, and Calibration Loop:** A continuous evaluation process in which simulated outputs are benchmarked against empirical evidence (e.g., shadow moiré warpage maps, digital image correlation, embedded strain-gauge telemetry, or targeted cross-sectional characterization) to quantify and reduce the model–reality discrepancy, establish confidence bounds, and prevent uncontrolled drift of calibrated parameters.

Although this architecture provides strong physical interpretability and mechanistic insight, the computational cost of repeatedly solving coupled PDEs at high resolution often imposes a prohibitive latency for in-line decision-making and real-time control. This limitation motivates reduced-order modeling and surrogate representations (discussed subsequently) that preserve essential physics while meeting operational timing constraints.

### B. Applications of Physics-Based Digital Twins

This subsection reviews representative physics-based Digital Twin implementations in 3D IC packaging, emphasizing how deterministic multiphysics modeling is operationalized for (i) defect localization and structural assurance, (ii) time-evolving material degradation, (iii) viscoelastic warpage prediction, and (iv) industrial-scale deployment. Table II summarizes the salient mechanisms, the functional role of the Twin, and the reported outcomes.

*1) Structural Assurance and Virtual Sample Preparation:* A distinctive use case for physics-based Twins is to *guide destructive failure analysis (FA)* by reducing the risk that sample preparation itself alters or destroys the defect of interest. Xu *et al.* [27] introduced a decision-support Twin aimed at high-quality sample preparation for advanced packages with buried defects.

- **Technical challenge:** In heterogeneous 3D stacks, defects may reside beneath opaque encapsulants and multilayer

dielectrics; unguided mechanical polishing and delayering can introduce preparation artifacts or remove the defect before it is observed.
- **Approach:** The framework couples terahertz time-domain spectroscopy (THz-TDS) with finite-element modeling to reconstruct layer-resolved structural/density information and to simulate the mechanical response during slicing and polishing.
- **Outcome and implication:** By predicting stress and warpage evolution during preparation, the Twin can recommend an optimal sectioning plane and process parameters, reducing preparation-induced cracking and improving confidence that observed damage corresponds to the true root cause rather than laboratory-induced artifacts.

*2) Lifecycle Degradation Modeling: Time-Evolving Material States:* While static simulations typically assume invariant material properties (e.g., elastic modulus and coefficient of thermal expansion (CTE)), a closed-loop physics-based Twin must explicitly accommodate property drift driven by aging and environmental exposure. Inamdar *et al.* [28] demonstrated this principle by modeling thermo-oxidative degradation in epoxy molding compounds (EMC).

- **Methodology:** The authors adopted a core–shell representation motivated by diffusion-limited oxidation, in which an "oxidized skin" layer grows over time (e.g., during high-temperature aging at 150°C) and is assigned distinct stiffness and CTE relative to the pristine core. The Twin updates the effective geometry and constitutive parameters as aging proceeds.
- **Significance:** By iteratively calibrating model parameters against experimental warpage measurements, the Twin captured time-dependent curvature evolution, including a reported warpage sign reversal that would be missed under static, time-invariant assumptions. This illustrates the essential DT capability of synchronizing a deterministic model with evolving material state.

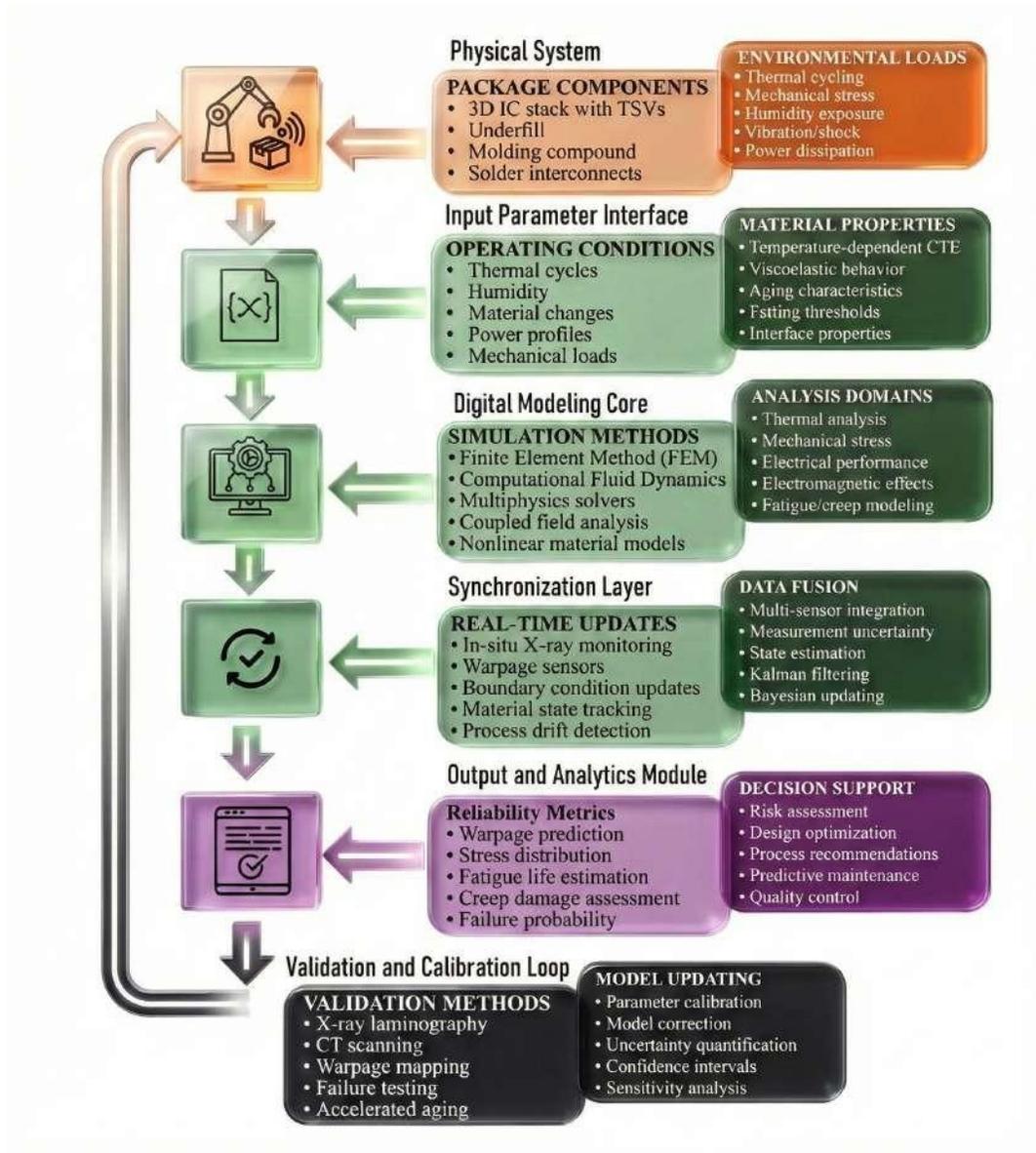

Fig. 7: Operational architecture of a Physics-Based Digital Twin. The framework is anchored by a deterministic Digital Modeling Core (employing FEM/CFD) that resolves multi-physics interactions based on explicit Input Parameters and material properties. Unlike static simulations, this architecture features a dynamic Synchronization Layer and a Validation/Calibration Loop that continuously update boundary conditions and constitutive parameters (e.g., via X-ray laminography or warpage mapping) to maintain fidelity against the evolving physical state of the 3D IC package.

*3) Viscoelastic Warpage Prediction:* Warpage in advanced memory stacks and molded packages is frequently governed by time-dependent polymer relaxation and temperature-dependent viscoelasticity, particularly during and after reflow. Kim *et al.* [29] developed a high-fidelity framework incorporating viscoelastic constitutive behavior (e.g., Prony-series relaxation) to predict transient and residual stress evolution.

- **Innovation:** In contrast to elastic models that assume instantaneous strain response, the viscoelastic formulation captures stress relaxation and rate effects in underfill/encapsulant materials during cooldown and sub-sequent dwell.

- **Outcome and reliability relevance:** By calibrating Prony-series parameters to dynamic mechanical analysis (DMA) data, the model achieved improved predictive agreement for both transient thermal gradients and long-term residual stresses, which are directly linked to delayed interconnect failures (e.g., cracks that initiate after prolonged field exposure rather than immediately post-assembly).

*4) Industrial Implementation Frameworks:* Scaling physics-based Twins beyond academic demonstrations

requires robust solver ecosystems, automation, and interoperability across EDA, multiphysics, and manufacturing data streams. Commercial platforms (e.g., Altair SimLab [30]) increasingly provide enabling infrastructure for DT-style workflows through parametric modeling, solver coupling, and automated analysis pipelines.

- **Role within the Twin:** Such platforms instantiate the *computational core* (Section IV-A), supporting automated coupling between electrical inputs (e.g., power maps) and thermo-mechanical solvers (e.g., stress/warpage), as well as parameterized geometry/material sweeps.
- **Enabling capability:** Automated design-of-experiments (DoE) and workflow orchestration allow systematic exploration of design/process variables (e.g., die thickness, bump pitch, underfill properties) and facilitate calibration against manufacturing feedback, improving model validity and uncertainty characterization at industrial scale.

### C. Data Assimilation and Model Updating Strategies

A critical differentiator between a static model and a Digital Twin is the latter's ability to autonomously correct its state using measurement data. In 3D IC packaging, where material properties drift due to aging and interfaces degrade under load, this requires rigorous Data Assimilation (DA) frameworks. Two primary strategies dominate the state-of-the-art:

*1) Bayesian Calibration (Slow-Time Updating):* For parameters that evolve over long horizons such as thermo-oxidative aging of molding compounds or creep in solder joints Bayesian inference provides a principled method to update model beliefs. By treating parameters (e.g., elastic modulus $E$, coefficient of thermal expansion $\alpha$) as probability distributions rather than fixed scalars, the Twin can ingest sparse periodic measurements (e.g., quarterly warpage scans or cross-sections) to narrow the uncertainty bounds of the constitutive model. This approach is particularly effective for calibrating viscoelastic Prony series coefficients, ensuring that FEA core accurately reflects the package's current relaxation behavior rather than its pristine "as-manufactured" state.

*2) Sequential State Estimation (Real-Time Updating):* For dynamic processes such as thermal management, Extended Kalman Filters (EKF) and Ensemble Kalman Filters (EnKF) offer low-latency state correction[16], [35]. In this scheme, the physics solver functions as the "process model," predicting the next thermal state based on power maps. Real-time deviations observed by embedded thermal diodes are then used to compute a Kalman gain, which instantaneously corrects the simulated temperature field. Crucially, these residuals can be mapped to specific physical drivers:

- **Thermal Mismatch → Contact Resistance:** Persistently higher measured temperatures at specific junctions can automatically trigger an increase in the modeled *thermal contact resistance* ($R_{th}$), serving as a proxy for interface delamination or underfill voiding.
- **Warpage Deviation → Effective Stiffness:** Discrepancies between predicted and measured warpage curvature

(e.g., via in-line profilometry) can be assimilated to update the *effective stiffness* of the redistribution layer (RDL), capturing the impact of latent cracking or layer decohesion.

### D. Insights

Physics-based DTs remain the cornerstone of structural and reliability analysis in 3D IC packaging because they are grounded in first principles of mechanics, thermodynamics, and materials science. As demonstrated in the reviewed works [27], [28], [29], finite element (FE)-based models can capture complex, multi-physics phenomena such as EMC oxidation-induced property shifts, time-dependent viscoelastic creep, and stress redistribution during warpage evolution. These models are indispensable for lifecycle-aware simulations, where accurate representation of evolving material states and geometric deformations is necessary to assess long-term reliability risks such as delamination, solder fatigue, or die cracking.

Another strength of physics-based DTs is their mechanistic interpretability: unlike data-driven methods, they provide causal insight into how process parameters and material properties influence failure mechanisms. This makes them particularly effective in high-stakes scenarios such as failure analysis, sample preparation guidance, and root-cause determination, as shown in structural analysis frameworks that integrate multi-modal measurements with FE simulations [27]. Similarly, commercial implementations such as ALTAIR SIMLAB [30] demonstrate that when combined with automated workflows and parametric models, physics-based DTs can bridge virtual prototyping with physical validation, thereby supporting both design optimization and lifecycle prognosis.

Nevertheless, physics-based DTs are less suited for exploratory design tasks that require rapid iteration. Their reliance on high-resolution FEM solvers makes them computationally expensive, often requiring hours to days of runtime for full-package simulations. Moreover, scalability to large design spaces or dynamic operating conditions is limited, as model recalibration and re-simulation must be performed whenever material properties, boundary conditions, or package layouts change. Thus, while physics-based DTs are unmatched in fidelity and interpretability, their primary value lies in high-precision reliability assessment and failure analysis rather than real-time monitoring or adaptive process control.

### E. Critical Challenges and Research Gaps

Despite their strong mechanistic interpretability and predictive capabilities, *pure* physics-based Digital Twins remain difficult to deploy at scale in high-volume 3D IC manufacturing. Across the literature and industrial practice, three persistent

TABLE II: Summary of Physics-Based Digital Twin Architectures for 3D IC Packaging

| Ref. | Physics Domain | Key Mechanism | Digital Twin Role | Outcome |
|---|---|---|---|---|
| [27] | Structural mechanics | THz-TDS coupled with finite-element analysis | **Decision support:** guided delayering and sectioning for FA | Reduces preparation-induced artifacts; improves defect observability in FA |
| [28] | Thermo-chemical aging | Diffusion-limited oxidation via core–shell modeling | **Lifecycle tracking:** time-evolving property/geometry updates | Predicts warpage evolution and sign reversal driven by EMC aging |
| [29] | Viscoelasticity | Prony-series relaxation calibrated to DMA | **Constitutive modeling:** time-dependent stress/warpage prediction | Improved prediction of post-reflow residual stress and long-term stress states |
| [30] | Multiphysics platform | Parametric FEM workflows and solver interoperability | **Industrial enablement:** automated coupling and DoE pipelines | Supports scalable design-for-reliability studies and model calibration workflows |

TABLE III: Summary of Numerical Data from Relevant Papers

| Paper | Application | Key Metric Reported | Numerical Value | Context |
|---|---|---|---|---|
| [28] | Degradation prediction | Oxidation Layer Thickness Model Fit | $d = 8.7 \ \mu m + 2.1 \ \mu m/h^{0.5} \times t^{0.5}$ | Accurate EMC oxidation model for physics-based DT. |
| [31] | Defect Prediction | Data Resolution (Occupancy Grid) | $32 \times 32 \times 64$ | 3D data resolution for glue; 20μm scan step. |
| | | Model Size (Simulation Model) | Channels: 1 to 512, then 1. Spatial: $32 \times 32 \times 64$ to $6 \times 6 \times 22$ | Shows 3DCNN complexity and data transformation. |
| [27] | IC Packaging (Sample Preparation) | THz-TDS Accuracy | Down to several $\mu m$ | High multi-layer thickness accuracy for IC prep. |
| [32] | 3D structure detection | 3D Object Detection (mAP) | 0.96 | High performance in detecting 3D interconnects. |
| | | 3D Segmentation (Dice Score) | 0.92 | Excellent 3D feature segmentation overlap. |
| | | 3D Metrology (Avg. Error) | 2.1 $\mu m$ | Precise buried structure measurement. |
| [29] | Warpage | Temperature Cycle (DMA Test) | $25°C$ to $245°C$ at $0.4°C/min$ | DMA characterizes EMC viscoelasticity for FEA. |
| | | Glass Transition Temp. (EMC) | $\approx 105°C$ | Critical property affecting warpage. |
| [33] | 3D-IC Thermal Simulation | Speedup vs. Solver | 1000x to 300000x | DeepOHeat accelerates 3D IC thermal simulation. |
| [34] | RUL prediction | Diagnostic Accuracy | 85% | Semi-supervised DT model achieved 85% accuracy. |
| | | Training Samples | 24 of 36 | Data proportion for model development. |
| | | Failure Criterion | Resistance > 50% of initial in 0.2s | Specific numerical criterion for failure. |
| | | BPNN Regression (R) | $\approx 0.86$ (20G), $0.83$ (30G) | High consistency for BPNN output. |

limitations recur, each directly constraining closed-loop operation and broad adoption.

*1) Computational Latency Barrier:* The dominant obstacle is computational cost. Full-package multiphysics analysis must reconcile disparate length and time scales from millimeter-scale global warpage and board constraint effects to micrometer-scale stress concentrations in micro-bumps, TSV vicinities, and interfacial corners often under nonlinear constitutive behavior (e.g., viscoplastic creep, temperature-dependent properties, and contact/interface nonlinearity). Even with commercial workflow automation, resolving coupled nonlinear PDEs for a full 3D stack can require hours (or longer) per evaluation, which is fundamentally incompatible with the sub-minute cycle times and low-latency decision requirements of in-line monitoring and control. Consequently, physics-based Twins are frequently confined to offline studies (design verification and reliability qualification) rather than real-time manufacturing control.

*2) The "Static Model" Trap:* A second limitation is the difficulty of maintaining a time-evolving, continuously synchronized model. Many reliability-relevant phenomena in advanced packaging are intrinsically non-stationary for example, EMC oxidation, polymer cure progression, moisture uptake, and damage accumulation which implies that effective material and interface parameters (e.g., $E$, $\alpha$, thermal conductivity, interfacial contact resistance) must be updated as the package ages and as degradation progresses. In prevailing workflows, such updates are typically performed manually through sporadic recalibration campaigns, yielding a *static* or piecewise-static model rather than a continuously adapting Twin. Achieving the DT ideal requires automated data assimilation and parameter/state estimation pipelines that can ingest degradation evidence and update the governing model in near real time; implementing this capability with conventional FEA remains computationally burdensome and is not yet routine in production environments.

*3) Expertise Bottleneck:* Finally, physics-based Twins impose a substantial expertise requirement. Accurate deployment depends on nontrivial modeling choices including boundary-condition specification, mesh design and convergence verification, constitutive model selection, parameter identification, and uncertainty management that typically require specialist training and extensive domain experience. This "expertise bottleneck" inhibits operational scalability and limits accessibility for manufacturing engineers and line operators, thereby restricting the practical impact of physics-based Twins in fab settings where rapid interpretation and action are essential.

These limitations motivate the shift toward *hybrid* Digital Twins that preserve the interpretability of physics while leveraging data-driven surrogates to meet latency and automation requirements. Before defining such hybrid architectures, however, it is necessary to examine the measurement infrastructure that enables synchronization and closed-loop operation namely, the *sensing pillar* that serves as the Digital Twin's "nervous system."

## VI. THE DATA PILLAR: DATA DRIVEN STATISTICAL MODELING

As established in Section V, high-fidelity physics-based solvers often incur a latency cost that is incompatible with real-time monitoring and closed-loop control. The *data-driven* paradigm mitigates this constraint by shifting the dominant computational burden from run-time numerical solution to offline model identification and training. Rather than explicitly solving governing constitutive relations (e.g., Navier–Stokes or coupled thermo-mechanical PDEs), data-driven Digital Twins employ statistical learning to approximate the functional relationship between measured inputs (e.g., tool telemetry and sensor streams) and target outputs (e.g., quality, reliability, or performance metrics), thereby enabling rapid inference at deployment. The generalized pipeline of the Data Pillar is illustrated in Fig. 9.

In 3D IC packaging, data-driven methods are particularly valuable for addressing the *observability limit*. Critical internal features including hybrid-bond interfaces, buried interconnects, and TSV sidewalls are not directly accessible to in-line inspection or during field operation, necessitating inference of latent states from available measurements. This motivates *virtual metrology* (VM), wherein the Digital Twin estimates unmeasurable or sparsely measured quality indicators from equipment and process data. Accordingly, this section surveys state-of-the-art machine-learning-based Digital Twins for advanced packaging, organizing prior work by (i) their capability to predict latent quality and reliability metrics and (ii) their effectiveness in anomaly detection within high-dimensional, heterogeneous data streams.

### A. Architecture of the Data-Driven Pipeline

A data-driven Digital Twin approximates system behavior via statistical inference, thereby avoiding the run-time cost of repeatedly solving coupled first-principles models. When applied to 3D IC packaging, the pipeline must be engineered for two dominant constraints introduced by heterogeneous integration: (i) the *observability limit*, where critical internal interfaces are not directly measurable in-line, and (ii) *heterogeneous, high-dimensional data*, spanning multi-rate time-series telemetry and multi-modal imaging. In practice, state-of-the-art implementations can be organized into four coupled stages:

- **Data Acquisition Layer (Digital Thread):** This stage consolidates disparate data streams into a unified, time-aligned record of the process and the evolving package state. In a 3D IC context, inputs commonly include temporal telemetry from in-situ or embedded sensors (e.g., on-die thermal diodes, piezoresistive stress gauges, tool health signals) as well as spatial and volumetric measurements from metrology modalities (e.g., scanning acoustic microscopy (SAM) delamination maps and 3D X-ray/CT projections). Because these sources differ in sampling rate, fidelity, and representation, robust implementations increasingly employ standardized sensor and interface abstractions (e.g., IEEE 1451 where applicable) to reduce data-format fragmentation and to support scalable fusion across tools, lots, and product variants.

- **Preprocessing and Feature Engineering:** Industrial datasets are typically affected by missing values, outliers, drift, and modality-specific artifacts; these issues are amplified in stacked assemblies where small misalignments can dominate downstream predictions. Consequently, preprocessing extends beyond denoising and normalization to include multi-modal synchronization, sensor validation, and geometry-aware alignment. Representative packaging-specific operations include image registration (e.g., aligning X-ray/CT volumes to CAD/layout coordinates), segmentation of interfaces and voids, and dimensionality reduction to mitigate the curse of dimensionality. Feature construction is most effective when it elevates physically meaningful descriptors such as TSV density and layout context, micro-bump offset distributions, warpage curvature profiles, and reflow thermal-history summaries while suppressing nuisance variability and measurement noise.

- **Learning Engine:** The learning stage maps acquired features to latent quality/reliability states and/or perfor-

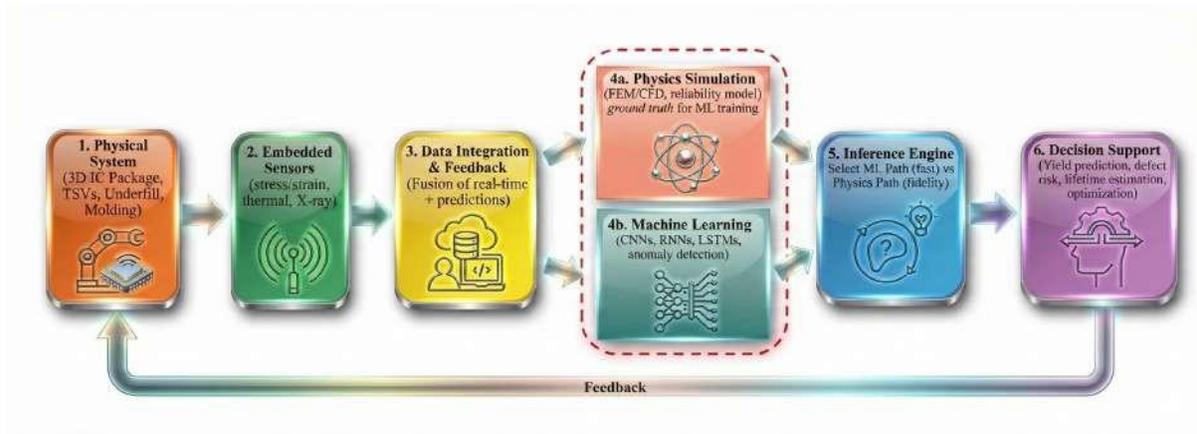

Fig. 8: Operational workflow of the Hybrid Digital Twin paradigm. The framework ingests multi-modal data from the Physical System and Embedded Sensors to drive a dual-branch modeling core (highlighted in red). In this architecture, Physics Simulations (4a) provide high-fidelity ground truth to train and validate Machine Learning models (4b). An Inference Engine (5) functions as an orchestrator, dynamically selecting the optimal path—leveraging ML for low-latency detection or Physics for high-fidelity resolution—to enable risk-aware Decision Support and closed-loop process optimization.

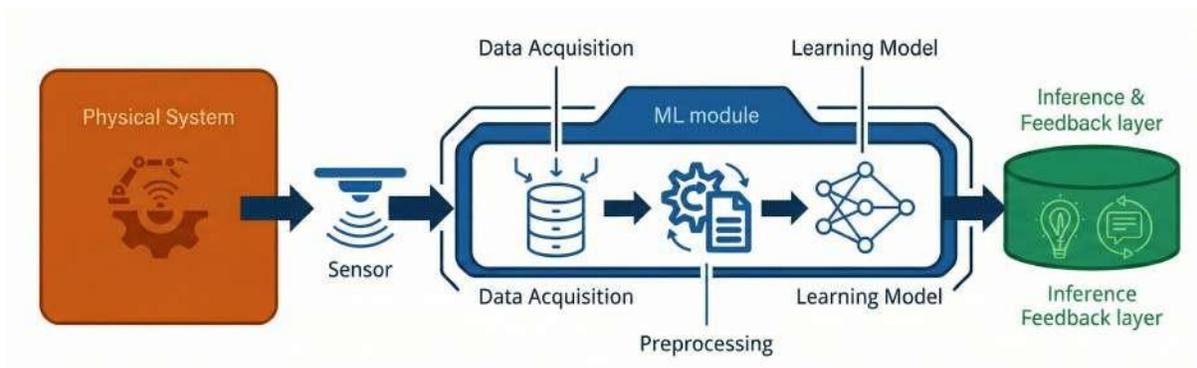

Fig. 9: Operational architecture of the Data-Driven Digital Twin. The pipeline illustrates the transformation of raw observability into actionable insight. Telemetry from the Physical System is captured via Sensors and routed to the ML Module, where it undergoes Data Acquisition and Preprocessing to extract meaningful features. The Learning Model (e.g., neural networks) then executes statistical inference to approximate system behavior, delivering real-time predictions via the Inference Feedback Layer to enable closed-loop control without the latency of physics-based solvers.

mance metrics, with model topology selected to match data structure and the targeted failure mechanism. Common choices include: (i) *convolutional neural networks (CNNs)* (including 3D CNN variants) for spatial defect recognition in SAM images and volumetric X-ray/CT data; (ii) *recurrent architectures* such as long short-term memory (LSTM) or gated recurrent units (GRUs) for history-dependent degradation modeling (e.g., fatigue accumulation under thermal cycling); and (iii) *Gaussian process regression (GPR)* and related probabilistic models for virtual metrology, where calibrated uncertainty estimates are essential for risk-aware decision-making rather than point prediction alone.

- **Inference and Feedback Layer:** The deployment stage produces low-latency estimates of otherwise unmeasurable metrics (e.g., probability of TSV cracking, likeli-hood of interfacial delamination, residual life under a given workload) and, when integrated into manufacturing, translates these estimates into actionable control or screening decisions. To be operationally relevant, inference is often executed at or near the tool/edge to meet cycle-time constraints, enabling timely parameter adaptation (e.g., reflow profile adjustment, process-window tightening, routing to additional inspection) before defects become latent escapes.

### B. Virtual Metrology: Inferring the Unmeasurable

In the 3D IC manufacturing loop, the *observability limit* implies that several critical quality attributes (CQAs)—including TSV etch depth, sidewall morphology, and hybrid-bond inter-face integrity—cannot be measured in-line at wafer or package scale without destructive analysis. Virtual Metrology (VM) addresses this limitation by learning surrogate regressors that infer metrology outcomes from equipment sensor data (ESD), tool telemetry, and process parameters, thereby enabling low-latency estimation of otherwise latent states and supporting run-to-run control.

TABLE IV: Summary of enabling ML methodologies for 3D IC Digital Twins, categorized by deployment context and latency requirements.

| Ref. | Modeling Paradigm | Target Application | Input Data | DT Output | Key Advantage | Deployment Setting | Latency Scale |
|------|-------------------|--------------------|------------|-----------|---------------|--------------------|--------------|
| [36] | Dual-phase (stepwise regression + BP neural network) | Virtual metrology (etch depth) | Tool telemetry (bias, pressure, flow) | Predicted process quality | Automated feature selection; high accuracy on tabular data | Inline (run-to-run) | Seconds |
| [32] | 3D CNN (computer vision) | Buried-structure metrology | Volumetric X-ray CT scans | Digital replica of as-built geometry | Non-destructive measurement of hidden micro-bumps | Inline metrology (post-bond) | Minutes (scan + reconstruction) |
| [33] | DeepOHeat (physics-informed) | Full-chip thermal management | Power maps; boundary conditions | Real-time temperature fields | $10^5\times$ speedup vs. FEM; generalizable | Design exploration & runtime control | Milliseconds (inference) |
| [34] | Ensemble learning (data-driven) | Interconnect prognostics | In-situ strain & temperature sensors | Remaining useful life (RUL) | Captures history-dependent degradation | Field operation (*in situ*) | Seconds (real-time update) |
| [31] | 3D CNN (geometric deep learning) | Bonding defect prediction | Point clouds / inspection logs | Defect classification | Predictive filtering for EDA simulations | Inline process control | Seconds (per unit) |

*1) Process Quality Prediction:* Cheng *et al.* [36] established the foundational dual-phase VM workflow—combining stepwise regression (SR) and back-propagation neural networks (BPNN)—which is widely applied to predict TSV etch depth and sidewall profiles from multivariate tool telemetry (e.g., chamber pressure, RF bias power, and gas flow rates). From a Digital Twin perspective, such VM capability is essential for shifting quality assessment from post-process verification (e.g., destructive cross-section SEM) to in-line, predictive monitoring: deviations between predicted and target etch outcomes can be interpreted as early indicators of process drift, enabling timely corrective action through recipe adjustment or run-to-run feedback. At the same time, the operational robustness of VM hinges on addressing tool aging, chamber-to-chamber variability, and distribution shift—factors that can degrade prediction fidelity unless uncertainty-aware learning and continuous recalibration are incorporated.

*2) Prognostics and Remaining Useful Life Estimation:* Beyond process monitoring, data-driven twins play a central role in prognostics and health management (PHM) by linking sensed operational signatures to degradation state and remaining useful life (RUL). A representative study on board-level interconnect reliability [34] employed in-situ strain gauges and temperature sensors to track solder joint degradation under thermal cycling, mapping measured hysteresis behavior to a damage-accumulation surrogate. The key methodological value for 3D IC packaging is the ability to learn degradation signatures under *non-ideal* and potentially stochastic loading histories—an advantage over canonical physics-of-failure rela-tions (e.g., Coffin–Manson) that often assume simplified cyclic conditions. When embedded within a DT loop, continuous updates of RUL based on realized usage history enable package-specific (i.e., personalized) reliability profiling, aligning with the DT definition in Section II by coupling measurement-driven state estimation with predictive decision support.

### C. Computer Vision for Buried-Structure Observability

While VM is frequently instantiated on time-series telemetry, advanced packaging also generates large-scale volumetric datasets from non-destructive testing (NDT), including 3D X-ray microscopy and acoustic imaging. In ultra-dense stacks, manual inspection of volumetric scans across thousands to millions of interconnect features (e.g., micro-bumps) is not tractable, motivating machine-learning-based computer vision to automate buried-structure observability and to convert imaging data into DT-ready state variables.

*1) Volumetric Reconstruction and Metrology:* Pahwa *et al.* [32] introduced a workflow for buried-structure characterization in dense metallic assemblies, where conventional X-ray microscopy is challenged by noise, beam-hardening, and reconstruction artifacts. Their approach combined detection and segmentation components tailored to 3D volumes:

- **Slice-and-fuse detection** to localize interconnect features within the reconstructed volume.
- **Anisotropic 3D CNN-based segmentation** to isolate individual micro-bumps and extract geometric attributes (e.g., diameter and standoff height) for quantitative metrology.

The reported performance (e.g., mean average precision of 0.96 and sub-3 $\mu$m metrology accuracy) enables a DT to transition from an *as-designed* representation (idealized CAD geometry) to an *as-built* digital replica of the manufactured structure. This distinction is consequential for downstream multi-physics analysis: mechanical and thermal simulations initialized with measured geometry can substantially improve predictive fidelity for warpage, stress concentration, and interface risk assessment relative to simulations driven by nominal design assumptions.

*2) Defect Prediction from Point-Cloud Representations:* Extending beyond reconstruction, Dimitriou *et al.* [31] proposed a 3D-CNN-based framework for defect prediction in conductive adhesive depositions using high-resolution point-cloud representations. By learning geometric precursors to electrical failure (e.g., insufficient volume and shape distortion), the model supports early classification of defect states prior to catastrophic outcomes. Importantly, the classification output is coupled to a downstream "simulation network" intended to extrapolate structural evolution, introducing a forward-looking capability in which current topology informs the likelihood and trajectory of future defect formation. In a DT setting, this coupling provides a practical pathway toward preventive control: process parameters can be adjusted proactively—rather than reactively—when predicted defect risk exceeds a threshold, thereby reducing latent escapes and improving yield.

### D. Critical Insights on Data-Driven Architectures

Across the reviewed literature, three recurrent patterns characterize ML-enhanced Digital Twin implementations for microelectronics packaging, each mapping directly to the principal barriers identified in Section I (observability, latency, and data sparsity).

*1) Observability Breakthrough:* Imaging-driven Twins exemplified by coupling X-ray microscopy (XRM) with deep learning for buried-structure characterization [32] represent the most direct response to the manufacturing "blind spot." By enabling high-resolution digital reconstruction of internal interconnect features, such frameworks support automated defect localization and screening that would otherwise require destructive analysis. Their primary advantage is non-destructive access to internal geometry and defect states at (reported) sub-micron metrology accuracy; however, translation to high-volume manufacturing remains constrained by the combined throughput limits of volumetric imaging and the computational expense of reconstruction and 3D inference, particularly when full-stack coverage and tight cycle-time budgets are required.

*2) Operator Learning Shift:* Simulation acceleration via physics-informed operator learning (e.g., DeepOHeat [33]) reflects a conceptual transition from explicitly *solving* governing equations to *inferring* the solution operator. By learning mappings from boundary/initial conditions and geometry to field solutions, deep operator networks can replace iterative PDE solvers and yield reported speedups on the order of $10^{3}$–$10^{5}\times$ while approaching the accuracy of commercial finite-element tools. This direction is especially attractive for real-time thermal monitoring and control in 3D stacks; nevertheless, robustness and transferability remain central concerns, as generalization to unseen chiplet topologies, material stacks, and power maps depends strongly on the diversity and coverage of the training physics, which can reintroduce substantial data-generation cost.

*3) Prognostic Capability:* Reliability-oriented Twins that fuse in-situ sensing with prognostic modeling (e.g., solder-joint RUL estimation under thermal cycling [34]) provide a practical route toward condition-based maintenance and reliability-aware operation. Unlike static qualification methodologies, these approaches can encode the realized usage history of individual packages, enabling personalized degradation tracking and life prediction. However, they are fundamentally limited by the *data sparsity paradox*: long-horizon run-to-failure datasets for advanced 3D packages are scarce and expensive to obtain, complicating rigorous validation—particularly for early-life and rare-event failure modes where predictive value is most critical.

*4) Geometric Deep Learning for Volumetric Defect Prediction:* While 2D CNNs are effective for surface inspection, defects in 3D IC packaging often manifest as *volumetric* anomalies (e.g., voiding in die-attach adhesives or interconnect misalignment) that are more faithfully captured using 3D sensing modalities such as inspection point clouds. Dimitriou *et al.* [31] addressed this challenge by developing a 3D deep learning framework tailored to the die-attach process.

Their approach departs from conventional image-based inspection by converting raw point clouds into 3D occupancy grids (i.e., voxelized representations). These volumetric inputs are then processed by a 3D convolutional neural network (3D-CNN) to classify the quality of conductive-adhesive deposition. A key contribution of the framework is its dual-purpose role within a Digital Twin workflow:

- **Predictive filtering:** The model serves as a high-speed inference engine (on the order of seconds per unit) to flag defective dies inline, preventing downstream value-add operations on known-bad parts.
- **Simulation integration:** By reconstructing the 3D geometry of the adhesive deposit, the model can provide boundary conditions for downstream finite-element thermal analyses, effectively automating the "scan-to-simulation" pipeline.

Overall, this work suggests that shifting from 2D pixel-based learning to 3D voxel-based learning is important for capturing the spatial complexity of heterogeneous packaging interfaces in 3D IC Digital Twins.

Overall, ML-based Digital Twins complement mechanistic modeling by extending predictive capability into regimes where physics-based simulation is either too slow for operational use (e.g., real-time thermal management) or insufficiently specified to capture emergent phenomena (e.g., defect formation and evolution). At present, their impact is primarily bottlenecked by limited availability of representative labeled datasets and by the lack of seamless co-integration with electronic design automation (EDA) and manufacturing execution workflows, which restricts scalability, traceability, and closed-loop deployment.

## E. Challenges and research gaps

Despite their promise, ML-based digital twins for 3D IC packaging face several unresolved challenges, foremost among them the scarcity of high-quality, labeled datasets a problem exacerbated by three packaging-specific factors: firstly,Destructive Ground Truth. Unlike image classification tasks where labels are readily visible, "ground truth" for buried defects (e.g., micro-bump cracking or hybrid-bond voids) often requires destructive physical failure analysis (PFA) techniques such as focused ion beam (FIB) cross-sectioning; this destructive requirement fundamentally limits dataset size to only dozens or hundreds of samples, preventing the "big data" scale typically assumed by standard deep learning; secondly, Label Noise and Variability. Even when defects are labeled, inter-annotator variability can be substantial, and subjective interpretation of X-ray or SAM artifacts produces noisy labels where one engineer might classify a marginal interface as a "void" while another flags it as "process variation," thereby degrading model convergence; and thirdly, Multi-Modal Misregistration. Effective fusion requires aligning data from disparate coordinate systems such as 3D X-ray volumes, 2D SAM maps, and idealized CAD layouts, and even slight registration errors (e.g., due to warpage-induced distortion not present in CAD) create "alignment noise" that decouples sensor features from their physical labels, confusing spatial learning algorithms.

A second challenge is model generalization. Many ML models are tailored to specific packaging geometries, materials, or operating conditions, leading to performance degradation when deployed on unseen architectures (e.g., new TSV layouts or underfill chemistries). Physics-informed learning and transfer learning offer partial remedies but require further research to ensure domain adaptability.

Third, explainability remains a bottleneck. Deep neural networks used for defect detection, thermal field regression, or RUL prediction are often black boxes, providing little insight into causal degradation mechanisms. This lack of interpretability limits trust in safety-critical applications and complicates integration with physics-of-failure frameworks.

Fourth, seamless integration with industrial EDA workflows is still underdeveloped. ML-based DTs must interface with multiphysics solvers, CAD/EDA tools, and process control systems, which currently lack standardized protocols for hybrid modeling. This hinders real-time deployment in design and manufacturing pipelines.

Finally, uncertainty quantification and robust validation frameworks are underexplored. Without calibrated confidence estimates, ML predictions risk misalignment with physics constraints, undermining reliability when extrapolated beyond training distributions.

These gaps highlight key research directions: developing standardized open datasets for packaging reliability, advancing hybrid ML–physics integration strategies, designing interpretable ML architectures, and creating common interface protocols for DT–EDA interoperability. Addressing these challenges will be essential for transitioning ML-enhanced digital twins from proof-of-concept demonstrations to scalable, industrially deployable solutions in 3D IC packaging.

## VII. THE HYBRID PILLAR: UNIFIED ARCHITECTURE

While the Physics and Data pillars offer complementary strengths, neither is sufficient in isolation for the complexities of 3D IC packaging. The literature currently treats these as competing alternatives; however, we argue that the future lies in their convergence. Synthesizing the disparate methodologies reviewed in Sections IV and V, we propose a unified, six-layer Hybrid Digital Twin architecture (Fig. 8). It is important to distinguish that while the individual constituent technologies (e.g., FEM solvers, CNNs, sensors) are established in the literature, the specific hierarchical integration and orchestration framework presented here is a novel conceptual contribution of this review, designed to reconcile the conflicting demands of real-time latency and high-fidelity resolution.

The *hybrid* Digital Twin paradigm has therefore emerged as a convergent solution that couples inductive learning with deductive, first-principles structure to mitigate the latency–fidelity tension. For 3D IC packaging, this convergence is particularly consequential: the Twin must reason over coupled multi-physics loading and failure mechanisms (requiring physics consistency) while simultaneously adapting to time-varying process conditions and degradation (requiring data-driven calibration and rapid inference).

### A. Generalized Hybrid Architecture

Based on the reviewed literature, a robust hybrid Digital Twin architecture for 3D IC packaging can be organized into six tightly coupled layers (Fig. 8):

- **Digital Baseline Layer (As-Designed Reference):** Encodes the authoritative design intent, including package geometry, material stacks and constitutive descriptions, interconnect topology, and nominal boundary/operating conditions. This layer provides a common reference representation that supports both deterministic solvers and learning pipelines, and enables traceability between "as-designed," "as-manufactured," and "as-operated" states.
- **Physics Core (High-Fidelity Reference):** Executes deterministic multiphysics solvers (e.g., FEM/CFD/EM) to resolve phenomena such as viscoelastic warpage, thermo-mechanical stress evolution, and current-driven degradation (e.g., electromigration). In a hybrid configuration, this core is typically invoked selectively: it functions as a *reference generator* for synthetic training data, as a validator for surrogate outputs, and as a diagnostic tool for rare or high-risk operating regimes where uncertainty must be minimized.

- **Learning Core (Low-Latency Inference):** Provides fast surrogate inference for tasks requiring millisecond-to-second response, such as real-time thermal field estimation, latent defect risk prediction, and anomaly detection. The learning core is commonly trained on physics-core simulations (to obtain broad coverage) and then adapted or recalibrated using empirical telemetry and metrology to correct model-form mismatch and reflect real manufacturing variability.

- **Data Assimilation Layer (State/Parameter Estimation):** Serves as the synchronization mechanism between physical measurements and the virtual state. By fusing in-situ sensor streams (Section V) with model predictions via, for example, Kalman filtering, Bayesian updating, or variational inference this layer estimates latent states and updates uncertain parameters (e.g., effective thermal contact resistance, stiffness degradation) to counteract drift induced by aging and damage accumulation.

- **Orchestration Layer (Fidelity Management):** Implements supervisory logic that allocates computation across the physics and learning cores as a function of risk, uncertainty, and operational context. Routine monitoring is handled by surrogates; when confidence is low, anomalies are detected, or excursions occur, the orchestrator escalates to higher-fidelity simulation and/or triggers targeted measurement, thereby operationalizing an adaptive "speed versus fidelity" trade-off.

- **Decision Support and Control Interface:** Translates fused state estimates into actionable recommendations or closed-loop commands for manufacturing and operation. Depending on the deployment setting, actions may include recipe adjustment (e.g., reflow profile tuning), routing decisions (e.g., additional inspection or rework), reliability-aware throttling (e.g., DVFS/thermal limits), or safety interventions (e.g., triggering holds or line stops) to prevent defect formation or mitigate risk before failures manifest.

### B. Applications of Hybrid Digital Twins

To date, fully realized hybrid Digital Twin frameworks *specific* to 3D IC packaging remain comparatively nascent. Despite extensive activity in surrogate modeling, multiphysics simulation, and in-situ sensing, there is a notable scarcity of implementations that unify the physics and data pillars into a single, operationally closed-loop architecture. This gap signals a clear research opportunity: while the constituent technologies are increasingly mature in isolation, their reliable integration into cohesive hybrid Twins with traceable state synchronization, uncertainty management, and actionable feedback arguably constitutes a remaining "final frontier" for advanced packaging.

The following subsections examine two foundational contributions that, while not exhaustive, provide instructive architectural blueprints for the emerging paradigm: an industrial patent articulation of modular hybridization and a methodological review proposing a dual-branch prognostics structure. Table V summarizes their target domains and division of labor between physics and learning components.

*1) Industrial Architecture: Modular Hybridization:* The U.S. patent "Digital Twin Modeling of IC Packaging Structure" [37] outlines an industrially oriented blueprint for a scalable hybrid Twin centered on modular decomposition rather than monolithic end-to-end modeling.

- **Core innovation:** A *modular surrogate strategy* partitions the Twin into interdependent functional modules (e.g., geometric representation, analytic/physics evaluation, learning-based estimation, and decision logic), enabling composability, maintainability, and targeted validation.

- **Hybrid mechanism:** Learned surrogates are trained using data generated by a physics module, but critically, an analytic module functions as a physics gatekeeper by screening surrogate outputs against established reliability relations (e.g., fatigue and life models) prior to permitting control-relevant actions.

- **Implication:** This design directly addresses trust and safety concerns associated with black-box predictors by enforcing physically motivated admissibility constraints, thereby aligning hybrid inference with the assurance requirements of high-volume semiconductor manufacturing.

*2) Prognostics Blueprint: Physics-of-Degradation Coupled with Data Inference:* In a methodological review, Deng *et al.* [38] propose a dual-branch hybrid architecture for prognostics and health management (PHM) that explicitly separates causal degradation modeling from measurement-driven inference.

- **Branch 1 (Physics-of-Degradation, PoD):** A mechanistic branch models dominant failure mechanisms grounded in physical chemistry and materials science (e.g., oxidation, intermetallic growth, fatigue kinetics), providing causal structure and interpretability for life prediction.

- **Branch 2 (Data-Driven Inference):** A statistical branch (e.g., Bayesian networks and related probabilistic models) processes real-time sensor evidence to accommodate noise, detect anomalies, and capture variability not represented in the mechanistic abstraction (e.g., stochastic assembly-induced defects).

- **Convergence mechanism:** The framework fuses branches using confidence-weighted logic in which the mechanistic model dominates early-life prediction when limited device-specific data are available, while the data-driven branch gains influence as operational evidence accumulates and device-specific degradation trajectories become identifiable.

### C. Critical Analysis: The Hybrid Imperative

In 3D IC packaging, the hybrid Digital Twin paradigm arguably represents the most credible route for reconciling the fundamental tension between high-fidelity multiphysics modeling and the latency requirements of in-line decision-making.

TABLE V: Emerging Hybrid Digital Twin Frameworks (Representative Examples)

| Ref./Type | Target Domain | Hybrid Mechanism | Role of Physics | Role of ML / Data |
|---|---|---|---|---|
| [37] | IC packaging design/analysis | Modular surrogate architecture with analytic gating | Reference simulation; admissibility screening via reliability laws | Low-latency inference; accelerates thermal/warpage estimation under constraints |
| [38] | Microelectronics PHM | Dual-branch fusion (PoD + data inference) with confidence weighting | Causal degradation modeling (e.g., oxidation/fatigue kinetics) | Anomaly detection; uncertainty-aware updating and device-specific calibration |

As suggested by modular and dual-branch architectures in [37], [38], hybridization can preserve physics consistency for multi-scale interactions (e.g., TSV-induced stress concentration, warpage-driven interfacial damage, and underfill delamination) while simultaneously enabling data-driven capabilities essential for Prognostics and Health Management (PHM), such as anomaly detection and device-specific degradation tracking.

Nevertheless, translating these architectural blueprints into industrially deployable systems remains constrained by three implementation barriers:

1) **Calibration challenge:** Sustained predictive validity requires continuous model updating under non-stationary material and interface states. As properties evolve due to thermo-oxidative aging, polymer cure progression, creep, or damage accumulation, the hybrid Twin must autonomously recalibrate physics parameters and surrogate corrections via measurement assimilation, without reliance on manual expert intervention.

2) **Interoperability gap:** Effective hybrid operation depends on fusing heterogeneous data streams spanning tool-level telemetry, in-line metrology, and in-field monitoring. Without rigorous standardization of sensing interfaces and data exchange (e.g., IEEE 1451 and UCIe-aligned pathways as discussed in Section V), integration remains ad hoc, hindering scalable deployment, traceability, and cross-platform portability.

3) **Trust barrier:** Deploying ML surrogates within safety- and yield-critical manufacturing loops requires uncertainty-aware decision logic. The Twin must quantify epistemic and aleatoric uncertainty, detect out-of-distribution conditions, and explicitly trigger escalation pathways for example, reverting to high-fidelity physics simulation or requesting additional measurement when surrogate predictions are insufficiently reliable.

**Conclusion:** Despite these challenges, hybrid Digital Twins provide a principled foundation for lifecycle reliability tracking in advanced packaging. By embedding physics-informed surrogates within modular, standards-aligned workflows, the hybrid paradigm elevates Digital Twins from passive simulation artifacts to adaptive, closed-loop frameworks for reliability-aware monitoring, prediction, and control, thereby narrowing the gap between academic modeling advances and scalable industrial deployment.

### D. Implementation Trade-offs and Challenges

Although hybrid Digital Twins provide a compelling resolution to the latency–fidelity tension, their practical adoption introduces several nontrivial trade-offs that must be managed explicitly at the algorithmic and system levels.

*1) Consistency–Efficiency Dilemma:* A central challenge is coordinating deterministic physics solvers with stochastic, data-driven surrogates in a manner that preserves physical admissibility. Ensuring that surrogate outputs remain consistent with conservation principles (e.g., energy balance and mass continuity) and constitutive constraints (e.g., yield surfaces and monotonicity under loading) typically requires structured coupling mechanisms, such as physics-informed loss functions, constrained architectures, or residual-correction formulations anchored to mechanistic baselines. Absent such safeguards, the Twin can generate computationally efficient but physically implausible predictions for example, nonphysical temperature fields, negative absolute temperatures, or stress states that violate material admissibility thereby undermining trust and limiting deployment in reliability- and yield-critical settings.

*2) Dual-Maintenance Burden:* Hybrid Twins also impose a two-front maintenance requirement driven by non-stationarity in both physics and data:

- **Physics drift:** Reference models must be updated as effective material and interface properties evolve due to thermo-oxidative aging, viscoelastic relaxation, moisture effects, creep, or damage accumulation.

- **Data drift:** Learned components are susceptible to concept and distribution drift when operating conditions (e.g., power maps, workloads, environmental profiles) or product configurations deviate from the regimes represented in training data.

Consequently, deployment requires a closed-loop validation and recalibration framework in which field or line telemetry continuously informs updates to *both* mechanistic parameters and surrogate representations, ideally with uncertainty-aware triggers that determine when retraining, re-identification, or model refitting is necessary.

*3) Integration Overhead:* Finally, hybridization increases system-level complexity and integration cost. Co-simulation across electrical, thermal, and mechanical domains requires multi-rate synchronization: electrical dynamics evolve on nanosecond time scales, whereas thermal transients often unfold over seconds and mechanical relaxation can extend further. Reconciling these heterogeneous temporal resolutions while concurrently ingesting multi-modal, multi-rate measurements from disparate sensing and metrology sources (Section V) introduces additional orchestration, data-management, and computational overhead that can partially offset the raw inference speed gains provided by surrogates.

### E. Dynamic Fidelity Switching Policy

A defining feature of the Hybrid architecture is its ability to dynamically allocate computational resources based on confidence levels. We define a Fidelity Escalation Policy managed by the Orchestration Layer, which defaults to low-latency surrogates during normal operation but triggers high-fidelity solvers when specific risk criteria are met. The switching logic is governed by three primary triggers:

*1) Uncertainty Quantification (UQ) Threshold:* The primary trigger is statistical uncertainty. By utilizing Bayesian Neural Networks (BNN) or Dropout Monte Carlo in the surrogate layer, the Digital Twin outputs both a prediction $\hat{y}$ and a confidence interval $\sigma$:

$$\text{IF } \sigma_{\text{pred}} > \tau_{\text{risk}} \implies \text{TRIGGER\_FEM.} \tag{1}$$

For example, if a surrogate model predicting TSV stress encounters an input vector (e.g., a power profile) significantly different from its training distribution (out-of-distribution), the variance $\sigma$ spikes. The Orchestrator detects this low confidence and automatically dispatches a job to the cloud-based FEA solver to resolve the state accurately, accepting the latency penalty for the sake of safety.

*2) Physics Consistency Checks:* The second trigger is physical viability. The Orchestrator continuously monitors the residuals of conservation laws (e.g., energy conservation in thermal blocks):

$$\text{IF } |\nabla \cdot (k \nabla T) - Q| > \epsilon_{\text{phys}} \implies \text{TRIGGER\_FEM.} \tag{2}$$

If a data-driven model predicts a physically impossible state such as a "cold spot" next to an active logic core (negative heat flux) the system flags the anomaly as a hallucination and reverts to the physics engine for a ground-truth calculation.

*3) Sensor–Model Divergence:* The final trigger is the "reality gap." The Digital Twin compares its prediction against real-time sensor telemetry. If the error between the predicted warpage (from the surrogate) and the measured warpage (from piezoresistive sensors) exceeds a calibrated tolerance, it indicates that the surrogate's underlying assumptions (e.g., material stiffness) have drifted. This triggers not only a high-fidelity simulation but also a Model Calibration Routine (as described in Section IV-C) to update the surrogate's parameters.

**Summary:** In many cases, the limiting factors for hybrid Digital Twins are not conceptual but operational: achieving trustworthy performance under evolving packaging architectures depends on maintaining interoperability at scale and enforcing rigorous model governance. Progress therefore hinges on standardized interface protocols (e.g., IEEE 1451 and related ecosystem standards) together with robust uncertainty quantification (UQ) and out-of-distribution detection to ensure that hybrid Twins remain reliable as materials, processes, and workloads change.

## VIII. Comparative Analysis of The Pillars

To provide a structured evaluation of the modeling paradigms reviewed in Sections IV through VI, we present a comparative profile in Fig. 6. It is important to note that the scores assigned in these radar charts are qualitative and indicative, representing a synthesis of the reported performance metrics in the surveyed literature rather than a single quantitative experiment. To ensure consistency and reduce subjectivity, the evaluation is grounded in the 5-point scoring rubric defined in Table VI, where a score of 1 represents the least favorable capability (e.g., highest latency or highest data cost) and a score of 5 represents the most favorable state for Digital Twin implementation.

The multi-physics and multi-scale nature of 3D IC packaging has motivated a heterogeneous set of Digital Twin implementations. As discussed in Sections III–V, three dominant modeling paradigms have emerged: (i) data-driven (ML-centric), (ii) physics-based (FEM/FEA-centric), and (iii) hybrid (physics–data co-designed) approaches. No single paradigm is universally optimal; rather, each occupies a distinct operating point on the Pareto frontier defined by predictive fidelity, inference latency, and interpretability. This section provides a structured comparison intended to guide researchers and practitioners in selecting architectures that align with specific reliability objectives and deployment constraints.

### A. Evaluation Criteria

To enable a rigorous comparison across paradigms, we evaluate each approach along eight dimensions relevant to the semiconductor lifecycle (summarized visually via the radar charts in Fig. 10):

- **Mechanistic fidelity:** Degree of adherence to governing physics (e.g., conservation laws and constitutive admissibility), supporting physically consistent extrapolation beyond observed data.
- **Inference latency:** Wall-clock time to generate actionable outputs after receiving inputs; in-line control and screening often require sub-second, and in some settings sub-100 ms, response.
- **Interpretability and diagnosability:** Ability to trace predictions to physically meaningful drivers (e.g., localized stress concentrations or thermal gradients), which is essential for failure analysis and design iteration.
- **Data dependency:** Quantity, diversity, and labeling cost of data required to develop and initialize the model (including defect/failure labels).
- **Generalizability and transfer:** Capacity to adapt to new package geometries, chiplet layouts, materials, and process windows with minimal retraining or re-identification.
- **Robustness to drift:** Resilience under process drift, tool aging, workload shifts, and distribution change, including mechanisms for detecting out-of-distribution conditions.

TABLE VI: Qualitative scoring rubric for modeling paradigms (Fig. 6).

| Axis / Metric | Score 1 (Low Capability) | Score 5 (High Capability) |
|---|---|---|
| **Inference Speed** | **Offline / batch:** requires an HPC cluster; hours to days (e.g., full FEA). | **Real-time:** millisecond latency; suitable for edge/closed-loop control. |
| **Data Efficiency** | **Data-hungry:** requires massive labeled datasets ($> 10^4$ samples); fails without big data. | **Zero-shot:** pure physics derivation; requires only geometry/material properties (no training data). |
| **Generalizability** | **Overfitted:** valid only for a specific trained geometry; requires retraining for new designs. | **Universal:** applicable across diverse packages and boundary conditions without modification. |
| **Interpretability** | **Black box:** no visibility into causal mechanisms (e.g., deep neural networks). | **White box:** fully explicit constitutive laws; physically verifiable. |
| **Fidelity** | **Trend only:** qualitative agreement; significant error ($> 10\%$) vs. ground truth. | **Metrology grade:** matches physical validation within experimental error ($< 2\%$). |

- **Uncertainty quantification (UQ):** Ability to provide calibrated confidence measures and risk-aware outputs, enabling safe decision-making and escalation to higher-fidelity analysis when needed.
- **Development and deployment cost:** Required human expertise, computational resources, software/toolchain complexity, and integration burden (including interoperability with EDA/manufacturing workflows).

### B. Critical Trade-off Analysis

Table VII synthesizes typical strengths and limitations of each paradigm against the above criteria.

*1) ML-Centric Paradigm (Low-Latency Inference):* As illustrated in Fig. 10(a), ML-based Twins typically excel in inference speed and in leveraging high-dimensional heterogeneous data (e.g., equipment telemetry and imaging), making them well suited for virtual metrology (Section III) and high-throughput defect screening where first-principles modeling is intractable or poorly specified (e.g., plasma chemistry and complex process signatures). The principal limitation is reduced mechanistic fidelity: without explicit constraints, purely data-driven models can produce non-physical predictions under sparse supervision or distribution shift, and they may require frequent retraining or adaptation as products, tools, and operating regimes evolve.

*2) Physics-Centric Paradigm (Mechanistic Ground Truth):* Figure 10(b) illustrates the complementary profile of physics-based Twins. By explicitly enforcing governing equations and constitutive behavior, physics-centric approaches offer strong interpretability and are indispensable for design verification, stress/warpage analysis, and root-cause reasoning, where understanding *why* a failure occurs can be as important as predicting *when*. Their limiting factor is computational latency and scalability: high-fidelity multiphysics simulations can be too slow for in-line control and may require substantial expert effort for model setup, calibration, and verification.

*3) Hybrid Paradigm (Risk-Aware Balance):* Hybrid Twins (Fig. 10(c)) provide the most balanced performance envelope by combining fast surrogates with physics-consistent structure. Hybridization can occur via physics-informed learning (constraining ML with conservation/admissibility), surrogate acceleration (learning operator or reduced-order representations of FEM/CFD), or modular gating strategies (using physics to validate ML outputs). The primary downside is integration complexity: robust hybrid deployment requires coordinated model governance (calibration, drift management, UQ), multi-rate orchestration across domains, and cross-disciplinary expertise spanning mechanics, thermal analysis, and statistical learning.

### C. Summary of Key Contributions

To consolidate the heterogeneous methodologies surveyed in this review, Table VIII organizes representative prior work according to the modeling paradigms and enabling pillars developed in Sections III–VI. The resulting taxonomy highlights the complementary roles of each approach: ML-centric studies tend to dominate tasks demanding low-latency inference and enhanced observability (e.g., virtual metrology and imaging-based inspection); physics-centric studies provide mechanistic validity and causal interpretability for coupled thermo-mechanical and degradation phenomena (e.g., warpage evolution and EMC oxidation); and hybrid architectures are increasingly positioned as the preferred pathway toward system-level, lifecycle-aware Digital Twins that can jointly satisfy fidelity, adaptability, and operational constraints.

### D. Challenges and Critical Barriers

As summarized in Fig. 8, the path to fully autonomous 3D IC Digital Twins is obstructed by five distinct categories of technical and structural barriers.

*1) Data Scarcity and the Rare-Event Paradox:* A primary bottleneck is the "data sparsity paradox" inherent to semiconductor reliability. Unlike consumer internet domains where labeled data is abundant, 3D IC failure signatures (e.g., microbump cracks, TSV delamination) are extremely rare relative to nominal operation, creating severe class imbalance. Furthermore, generating "ground truth" often requires destructive Physical Failure Analysis (PFA), which fundamentally limits dataset size to dozens rather than millions of samples. This scarcity makes it difficult to train deep learning models without overfitting, necessitating reliance on synthetic data augmentation or physics-informed regularization, which introduce their own uncertainties.

TABLE VII: Strategic Comparison of Digital Twin Modeling Paradigms for 3D IC Packaging

| Dimension | ML-Centric (Data-Driven) | Physics-Centric (White Box) | Hybrid (Grey Box) |
|---|---|---|---|
| Primary advantage | Low-latency inference; strong performance on high-dimensional telemetry/imaging; scalable screening | High mechanistic consistency; strong diagnosability and extrapolation via governing laws | Balanced speed and fidelity via surrogates and physics constraints; supports risk-aware escalation |
| Primary limitation | Limited physical guarantees; vulnerable to distribution shift without drift/UQ mechanisms | High computational cost; difficult to meet in-line latency constraints at full-stack resolution | System integration complexity; requires coordinated calibration, UQ, and orchestration |
| Data dependency | Moderate-to-high (especially for labeled defects/failures); sensitive to coverage and bias | Low-to-moderate (geometry/materials dominate); still requires calibration/validation data | Moderate: physics reduces labeled-data burden; still needs telemetry for correction and drift tracking |
| Transfer to new designs | Often limited without adaptation; retraining/transfer learning commonly required | High: governing equations apply broadly; updates mainly via geometry/material changes | High when physics backbone is retained; surrogates can be adapted with smaller data |
| Typical use cases | Virtual metrology; anomaly detection; imaging-based defect classification | Design verification; stress/warpage analysis; root-cause analysis and what-if studies | Lifecycle prognostics (RUL); real-time process control with confidence gating; self-calibrating monitoring |

*2) Computational Bottlenecks and Cost:* High-fidelity multiphysics solvers (FEM/CFD) face a prohibitive "accuracy–latency" trade-off. Resolving the stress gradients in a 60,000-bump chiplet assembly requires mesh densities that take hours or days to solve on HPC clusters, rendering them unusable for real-time control. While Reduced-Order Models (ROMs) offer speedup, they often fail to capture nonlinear constitutive behaviors (e.g., solder creep) accurately under varying boundary conditions. This computational burden creates a resource strain, where the cost of training and maintaining high-fidelity surrogates ($) can outweigh the immediate yield benefits for cost-sensitive packaging lines.

*3) Real-Time Latency and Bandwidth Limits:* As highlighted in the "Real-time Issue" branch of Fig. 6, operational Digital Twins face strict timing constraints. In-line process control often demands inference within milliseconds to trigger tool interlocks. However, the transmission of raw high-bandwidth telemetry (e.g., 3D X-ray volumes or high-speed RF waveforms) to a centralized cloud Twin can saturate network bandwidth. This creates a "data gravity" problem where the sheer volume of sensor data exceeds the pipe's capacity, forcing a shift toward edge-computing architectures that are currently immature in backend packaging facilities.

*4) Lifecycle Discontinuity (Design vs. Operation):* Currently, Digital Twins are largely confined to the design stage ("digital prototypes"), with little continuity into manufacturing or field operation. There is a "silence" between the EDA environment where the ideal 3D stack is defined and the physical fab data systems. Consequently, the Twin used for reliability qualification is rarely the same entity used for process monitoring. This lack of end-to-end synchronization prevents the feedback loop where field failure data updates the design-stage physics models, breaking the promise of continuous lifecycle learning.

*5) The Standardization Gap:* Finally, the industry lacks unified protocols for data and model exchange. Design intent resides in proprietary EDA formats (GDSII/OASIS), simulation data in vendor-specific FEA binary files, and manufacturing logs in unstructured CSV/JSON streams. The absence of a common ontology (such as emerging IEEE 1451 or IPC-2581 standards) hinders interoperability, forcing engineers to build brittle, ad-hoc data pipelines. This fragmentation isolates the "Physics Pillar" from the "Data Pillar," making scalable, multi-vendor Digital Twin deployment nearly impossible.

*E. Future Directions*

*1) Generative AI and Synthetic Data for Augmentation:* To mitigate data scarcity and imbalance, an emerging direction is the use of generative modeling and synthetic data generation to enrich training distributions, particularly for rare defects and early-life failure signatures. Generative adversarial networks (GANs) and diffusion-based models can be used to synthesize realistic defect morphologies, imaging artifacts, and process anomalies, enabling pre-training and stress-testing of surrogates in data-poor regimes. When combined with physics-based simulation (for controllable ground truth) and uncertainty-aware filtering (to avoid propagating unrealistic samples), synthetic augmentation can improve robustness and reduce the dependence on costly labeled failure data prior to deployment on production tools.

*2) Standardized Digital Twin Frameworks and Semantic Interchange:* Scalable Digital Twin adoption will require industry-aligned standards for data exchange and semantic interoperability across design, manufacturing, and test domains. Future efforts should prioritize common interface formats and co-simulation standards (e.g., FMU/FMI where appropriate), alongside domain-specific ontologies that encode packaging structures, process context, and reliability states in machine-interpretable form. Interoperable, open tooling including reference implementations that bridge EDA abstractions with multiphysics models and manufacturing telemetry would reduce integration overhead, improve traceability, and lower the barrier to deploying hybrid Twins across heterogeneous vendor ecosystems.

TABLE VIII: Taxonomy of Representative Digital Twin Research for 3D IC Packaging

| Reference | Paradigm | Target Application | Key Innovation / Contribution |
|-----------|----------|--------------------|-------------------------------|
| [32] | ML-centric | Buried-structure metrology | Developed a 3D-CNN pipeline for non-destructive segmentation and geometric characterization of hidden micro-bumps from X-ray/CT data. |
| [33] | ML-centric (operator learning) | Full-chip thermal management | Proposed DeepOHeat (DeepONet) to learn a solution operator for thermal fields, reporting up to $10^5\times$ speedup over FEM for real-time thermal mapping. |
| [34] | ML-centric | Interconnect prognostics | Demonstrated data-driven RUL estimation using in-situ strain/temperature telemetry to learn degradation signatures. |
| [31] | ML-centric | Defect screening | Combined geometric deep learning with simulation-informed supervision to classify defect-prone bonding/adhesive configurations. |
| [29] | Physics-centric | Warpage prediction | Incorporated viscoelastic constitutive modeling (Prony-series relaxation) to capture time-dependent stress relaxation in molding/underfill materials. |
| [28] | Physics-centric | Lifecycle degradation | Introduced a time-evolving Twin using a core–shell model to represent diffusion-limited thermo-oxidative aging of EMC and its impact on warpage. |
| [27] | Physics-centric | Virtual sample preparation (FA) | Coupled sensing with FEM stress/warpage prediction to guide delayering and reduce preparation-induced artifacts in failure analysis. |
| [30] | Physics-centric | Industrial design platform | Provided parametric multiphysics workflows enabling automated design-of-experiments (DoE) and scalable design-for-reliability studies. |
| [37] | Hybrid | Industrial architecture | Proposed a modular "gatekeeper" architecture in which physics-based admissibility checks validate ML surrogate outputs prior to control actions. |
| [38] | Hybrid | Prognostics (PHM) | Presented a dual-branch strategy that fuses physics-of-degradation modeling with data-driven anomaly detection via confidence-weighted integration. |

*3) Cloud–Edge Continuum for Low-Latency Control:* As Digital Twins evolve from monitoring artifacts to closed-loop control systems, compute architecture becomes a first-order design consideration. A promising deployment model is a cloud–edge continuum in which computationally intensive workloads (e.g., large-scale simulation, surrogate training, and periodic model re-identification) execute in the cloud or centralized infrastructure, while latency-critical inference and control operate at the edge co-located with manufacturing equipment or embedded within system firmware. This division enables rapid response to process drift and reliability hazards while preserving the ability to perform periodic high-fidelity updates, thereby aligning operational latency constraints with lifecycle model governance.

*4) Data Scarcity and Observability:* A foundational barrier to robust ML-centric and hybrid Digital Twins is the limited availability of high-quality labeled datasets, particularly for rare and safety-relevant failure modes. In 3D IC packaging, critical structures such as through-silicon vias (TSVs) and micro-bumps are buried within stacked layers, so obtaining internal ground truth often requires slow, costly, or throughput-limited non-destructive inspection (e.g., X-ray microscopy/CT or scanning acoustic microscopy (SAM)) and, in many cases, destructive confirmation. As a result, available datasets are typically dominated by nominal "pass" samples with comparatively few defect instances (e.g., delamination, cracking, voiding), producing severe class imbalance and biased coverage. Beyond scarcity, multi-modal fusion poses an additional obstacle: combining multi-rate time-series telemetry with spatial/volumetric imaging demands precise synchronization, registration, and alignment (including coordinate-frame consistency with layout/CAD), capabilities that remain difficult to automate reliably in current manufacturing data pipelines.

*5) Multiphysics and Multi-Scale Complexity:* 3D heterogeneous integration introduces tightly coupled interactions across thermal, mechanical, and (often) electrical domains, spanning wide ranges of spatial and temporal scales. For example, accurate prediction of thermo-mechanical hysteresis and fatigue accumulation may require consistent representation of nanoscale and microscale mechanisms (e.g., intermetallic compound (IMC) formation and microstructural evolution) together with mesoscale interconnect behavior and package-scale phenomena such as wafer-level warpage. High-fidelity first-principles solvers (e.g., FEM/FEA and CFD where applicable) are generally computationally prohibitive for such multi-scale, nonlinear problems under real-time constraints. Developing reduced-order models (ROMs) and surrogates that preserve fidelity across these disparate scales without

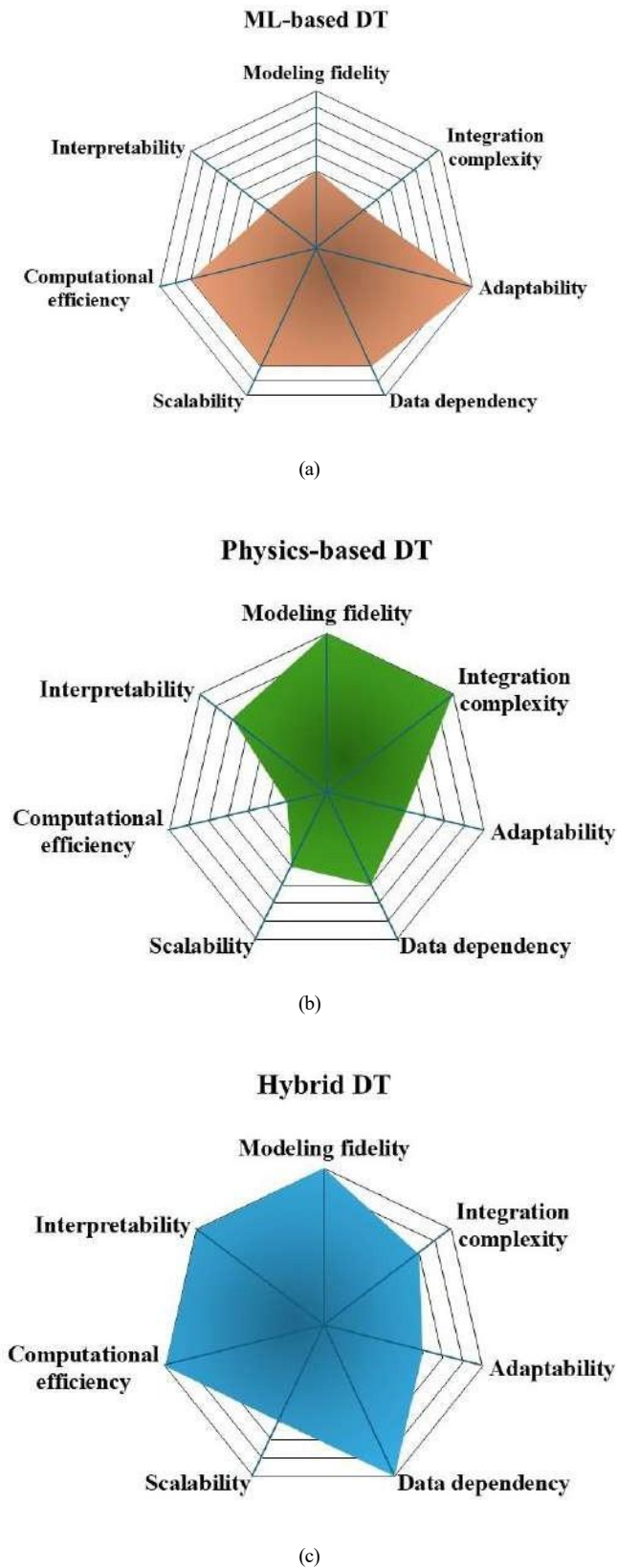

ML-based DT

(a)

Physics-based DT

(b)

Hybrid DT

(c)

Fig. 10: Indicative radar-chart comparison of modeling paradigms: ML-centric, physics-centric, and hybrid Digital Twins. Hybrid approaches typically expand the feasible operating envelope by mitigating the latency of physics-only solvers and the non-physical extrapolation risk of unconstrained ML models

oversimplifying interface behavior, contact nonlinearity, or history-dependent constitutive effects remains a key modeling bottleneck for deployable Twins.

*6) Interoperability and Standardization Gaps:* Digital Twin realization is further hindered by a fragmented semiconductor toolchain: design intent and layout are encapsulated in EDA environments; reliability physics and multiphysics analysis are performed in FEA/CFD platforms; and manufacturing telemetry is captured in proprietary equipment ecosystems (often mediated by SECS/GEM and vendor-specific schemas). The absence of a unified ontology and broadly adopted exchange standards tailored to IC packaging (including consistent semantics for structures, materials, process context, and reliability state) forces reliance on ad hoc scripts and brittle point-to-point interfaces. This siloed architecture undermines traceability, complicates validation, and materially increases the integration effort required to build lifecycle-aware Twins that span design, assembly, and field operation.

*7) Real-Time Latency versus Fidelity Trade-off:* For a Digital Twin to support active process control or runtime mitigation, inference must execute within manufacturing takt time (often milliseconds to seconds). Yet high-fidelity physics-based models can require minutes to hours to converge, especially under coupled nonlinear constitutive behavior. Although pure ML surrogates can meet strict latency budgets, their deployment in yield- and safety-critical loops is limited by generalization risk under distribution shift and by insufficient physical guarantees outside the training envelope. Achieving operationally reliable "hard" real-time performance therefore requires edge-compatible architectures that couple lightweight inference at the tool/edge with periodic high-fidelity recalibration and retraining in centralized infrastructure (cloud or on-prem), a compute-governance paradigm that is still maturing for semiconductor manufacturing.

*8) Intellectual Property (IP) and Security:* Digital Twins aggregate sensitive artifacts including design intent, layout-derived structural features, material parameters, process recipes, and in-line telemetry making them high-value targets for IP theft and integrity attacks. In disaggregated supply chains where chiplets, interposers, substrates, and assembly/test are distributed across multiple vendors, sharing a full-fidelity Twin can expose proprietary trade secrets and create unacceptable risk. The limited availability of secure, privacy-preserving collaboration mechanisms (e.g., federated learning, secure enclaves, encrypted model exchange, and policy-governed data minimization) therefore constrains cross-organizational adoption, particularly among foundries and OSAT providers, and motivates research on DT architectures that enable utility without full information disclosure.

*9) Lack of Standardized Architecture:* Despite the promise of hybrid Digital Twins, the absence of unified architectural standards continues to constrain interoperability and scalability across the semiconductor ecosystem. Reported implementations often adopt incompatible modeling granularities (die, package, board, system), inconsistent data abstractions, and bespoke interface protocols, effectively creating silos that decouple EDA-driven design intent from manufacturing execution systems and in-situ sensing platforms. This fragmentation inhibits formation of a cohesive tool-and-data ecosystem, slows industrial deployment, and complicates cross-domain exchange of state variables, uncertainties, and control-relevant outputs.

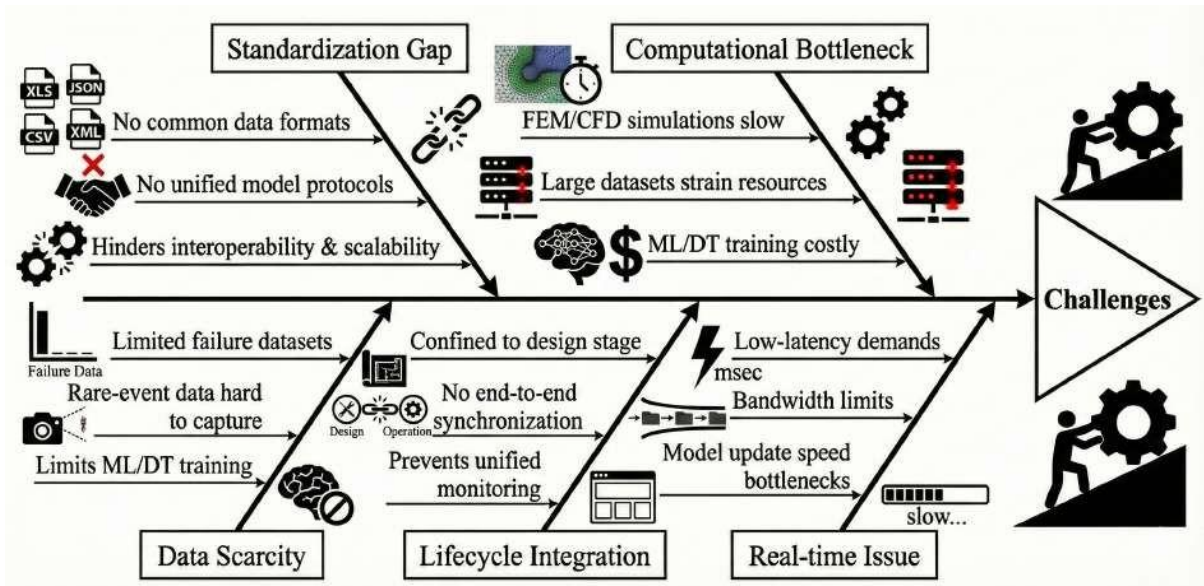

Fig. 11: Cause-and-effect diagram summarizing the primary barriers to Digital Twin adoption in 3D IC packaging. The challenges are categorized into five critical domains: Standardization Gap (lack of unified data/model protocols), Computational Bottleneck (latency of high-fidelity solvers), Data Scarcity (limited availability of rare-failure datasets), Real-time Issues (bandwidth and inference speed limits), and Lifecycle Integration (discontinuity between design and operation phases).

*10) Data Scarcity and Data Quality:* ML-centric and hybrid Twins are fundamentally limited by access to extensive, high-fidelity datasets, yet data acquisition in 3D IC packaging is structurally constrained. Degradation and failure events are intrinsically rare due to long device lifetimes and high baseline reliability, yielding severe class imbalance and limited coverage of corner-case behaviors. Moreover, non-destructive internal imaging (e.g., X-ray/CT and SAM) is capital-intensive, throughput-limited, and often sensitive to package-specific materials and geometries, while the resulting measurements can be noisy and artifact-prone. Finally, reliance on manual defect labeling and expert adjudication restricts dataset scale and diversity, reducing model transferability across architectures, process nodes, and product families.

*11) Computational Bottlenecks in Physics-Based Systems:* While physics-based Twins provide strong mechanistic fidelity, their practicality is bounded by the computational cost of finite-element and coupled multiphysics solvers. Full-package simulations with nonlinear constitutive behavior and multiscale resolution can require hours to days per run, imposing a latency barrier that precludes real-time use cases such as in-line monitoring, closed-loop control, or rapid screening. The same computational burden also restricts design-space exploration during prototyping, limiting the number of candidate geometries, materials, and process conditions that can be evaluated within development timelines.

*12) Model Interpretability and Explainability:* Purely data-driven Twins frequently operate as opaque predictors, generating outputs without transparent causal attribution. In yield- and safety-critical contexts including defect screening, process excursion diagnosis, and remaining useful life (RUL) estimation limited interpretability impedes trust, complicates qualification, and reduces actionable insight for remediation. Although explainable AI (XAI) provides general techniques for attribution and uncertainty reasoning, packaging-specific reliability applications remain under-served: methods that align explanations with physics-of-failure concepts (e.g., stress hotspots, interfacial energy release, fatigue drivers) are not yet mature or widely adopted.

*13) Integration with Real-Time Manufacturing Systems:* Closing the loop between a Digital Twin and the physical manufacturing line requires low-latency sensing, deterministic data transmission, reliable time synchronization, and adaptive model updating under tight cycle-time constraints. Many packaging facilities lack the necessary IT/OT integration including unified data buses, standardized telemetry interfaces, and robust edge compute to support continuous bidirectional flow between equipment, metrology, and DT services. As a result, real-time feedback control remains technically challenging, and many deployments remain limited to offline analytics rather than online process optimization.

*14) Lifecycle Synchronization:* Most current implementations target isolated lifecycle stages (e.g., design verification, prototyping, or reliability testing) rather than sustaining a continuous digital thread across design, fabrication, assembly, and field operation. Without a lifecycle-aware DT architecture, insights obtained from manufacturing excursions and in-field degradation cannot be propagated systematically back into design rules and process windows. This discontinuity limits holistic reliability management, inhibits systematic learning across product generations, and reduces the ability to optimize long-term performance through closed-loop, lifecycle-spanning governance.

*F. Future Research Directions*

Addressing the foregoing challenges will require coordinated advances spanning modeling, data infrastructure, compute architectures, and governance. The research directions outlined below (summarized with technical opportunities in Table IX) represent high-leverage steps toward mature, scalable Digital Twin deployment for 3D IC packaging.

*1) Modular, Open, and Standards-Aligned DT Frameworks:* A pressing industrial need is to move beyond one-off, ad hoc implementations toward modular, composable, and standards-aligned DT ecosystems. Next-generation frameworks should provide interoperable abstractions across EDA design environments, multiphysics solvers (FEA/CFD/EM), and heterogeneous metrology/sensing platforms, with explicit support for traceability across "as-designed," "as-built," and "as-operated" states. Progress will likely depend on converging toward shared ontologies for model/data exchange (e.g., FMI/FMU-style interchange where appropriate) and harmonized communication protocols that dismantle current data silos and enable vendor-agnostic adoption across the disaggregated semiconductor supply chain.

*2) Synthetic Data Generation and Augmentation for Rare Events:* To mitigate the scarcity of labeled failures and rare degradation signatures, synthetic data generation is poised to become a foundational enabler. Generative adversarial networks (GANs), variational autoencoders (VAEs), diffusion models, and multi-fidelity physics simulators can be leveraged to expand training distributions for low-probability defect morphologies and degradation trajectories. When combined with physics-based admissibility filters and uncertainty-aware selection, synthetic augmentation can reduce reliance on capital-intensive experiments, stabilize ML training under class imbalance, and improve generalization across packaging architectures, materials, and process nodes.

*3) Generative-AI-Augmented Digital Twins with Physics Guardrails:* Generative AI (GenAI) introduces the possibility of constructing or completing Twin artifacts from sparse inputs for example, proposing boundary conditions, geometry hypotheses, or plausible degradation scenarios using diffusion models and large language models (LLMs). However, unconstrained GenAI risks producing physically invalid or non-executable configurations. A promising frontier is the development of "fusion intelligence" frameworks that couple GenAI with physics-AI (PhyAI) guardrails, enforcing mathematical consistency and physical realizability via constraints, differentiable solvers, or learned admissibility checks [39].

*4) Self-Calibrating and Self-Validating Digital Twins:* Sustained DT fidelity under process drift and material aging demands a shift from static calibration to continuous, automated alignment. Analogous to emerging CFD–sensor fusion paradigms [?], 3D IC Twins can assimilate real-time evidence from warpage mapping, X-ray laminography, and embedded thermal/strain sensors to update latent states and uncertain parameters (e.g., effective CTE, modulus evolution, thermal contact resistance). Bayesian calibration, ensemble Kalman filtering, and related data-assimilation methods provide principled routes to quantify uncertainty while maintaining synchronization as the package evolves.

*5) Cloud–Edge Collaborative DT Architectures:* Exclusive reliance on centralized/cloud compute can introduce unacceptable latency for in-line control and edge decision-making. Scalable deployment will therefore require a cloud–edge continuum in which computationally intensive tasks (e.g., high-fidelity simulation, surrogate training, and global uncertainty quantification) execute centrally, while latency-critical inference runs at the tool/edge. Dynamic task-offloading and scheduling algorithms [40] will be essential to minimize bandwidth and ensure responsiveness within manufacturing takt times.

*6) Next-Generation Physics-Informed Machine Learning:* Beyond canonical physics-informed neural networks (PINNs), emerging architectures such as physics-encoded residual blocks [41] suggest that embedding conservation laws and constitutive structure directly within network layers can improve stability and data efficiency. For 3D IC packaging, such approaches offer a route to capture coupled heat transport and thermo-mechanical response with substantially reduced labeled-data requirements compared to purely data-driven models, while preserving interpretability and physical admissibility.

*7) Edge-Compatible DT Implementation and Model Compression:* Real-world fab deployment requires inference pipelines that operate reliably on resource-constrained edge devices. Research on quantization, pruning, knowledge distillation, and hardware-aware neural architecture search can enable sophisticated defect screening and warpage/thermal risk estimation to run locally [42]. Executing time-critical inference at the edge while synchronizing asynchronously with cloud-based model governance improves fault tolerance, reduces network dependence, and supports decentralized real-time decisions without saturating telemetry bandwidth.

*8) Digital Twins for Additive Manufacturing-Enabled Packaging:* Additive manufacturing (AM) is increasingly explored for redistribution layers (RDLs) and high-density interconnect formation. DT methodologies developed for AM provide transferable paradigms for monitoring and controlling geometry evolution and process-induced defects. For example, voxel-based physics-informed Twins [43] have coupled extrusion dynamics with evolving geometry to predict layer morphology, while self-correcting DTs integrating LSTM predictors with Bayesian calibration [44] demonstrate dynamic process optimization. Translating these concepts to semiconductor packaging could enable layer-wise monitoring of printed interconnects and closed-loop mitigation of warpage and defect

TABLE IX: Summary of critical research opportunities and technical enablers for next-generation 3D IC Digital Twins.

| Research Domain | Current Limitation | Technical Opportunity | Key Enablers | Expected Impact |
|---|---|---|---|---|
| **Physics Modeling** | **Computational latency:** high-fidelity FEA is too slow for real-time control. | **Real-time surrogates:** develop physics-encoded ML models that respect conservation laws without iterative solving. | PINNs, DeepONet | Enable ms-scale thermal/stress inference for active throttling. |
| **Data Engineering** | **Data sparsity:** critical defects (voids, cracks) are rare events; labeling requires destructive PFA. | **Synthetic augmentation:** use generative AI to create realistic "virtual defects" to train classifiers. | GANs, diffusion models | Overcome the "small data" paradox in reliability training. |
| **Sensing & Metrology** | **Observability limit:** buried interfaces (hybrid bonds) are invisible to optical/IR inspection. | **Deep in-situ telemetry:** integrate nanoscale sensors directly into the interconnect stack (BEOL). | Graphene nanoribbons (GNR), carbon nanotubes (CNT) | Granular visibility into local hotspots and stress concentrations. |
| **Standardization** | **Fragmentation:** disconnected tools (EDA vs. fab) and proprietary data formats. | **Universal interoperability:** establish unified ontologies for seamless data exchange across the lifecycle. | IEEE 1451 (smart transducers), UCIe (chiplet interface) | "Plug-and-play" Digital Twin deployment across multi-vendor lines. |
| **Control Systems** | **Open-loop monitoring:** Digital Shadows only visualize data; they do not act. | **Autonomous correction:** close the loop with reinforcement learning agents that adjust process parameters. | DRL, edge computing | Self-healing manufacturing lines with zero-touch optimization. |

formation.

*9) Integration of Emerging Materials, Interconnects, and Cyber–Physical Infrastructure:* Future DT frameworks must anticipate post-Moore materials and architectures that introduce non-classical transport and reliability behavior. Novel interconnect and device technologies including graphene nanoribbon FETs and carbon nanotube bundles require models that capture atypical electrical/thermal transport characteristics [22], [45]. In parallel, cyber–physical infrastructures leveraging distributed machine learning and low-latency analytics [46], [5] can support real-time monitoring and control at scale. A long-term objective is hybrid DT software that bridges nanoscale physics fidelity with industrial-grade scalability and maintainability.

*10) Closed-Loop Control and DT–Actuator Co-Design:* The ultimate evolution of Digital Twins is a transition from passive diagnosis to active prescription and intervention. Advancing toward this goal requires robust DT–actuator interfaces and control policies that translate state estimates and risk metrics into reliable actions, such as adaptive thermal throttling, stress-aware workload scheduling, and real-time process parameter tuning. Research emphasis should include stability guarantees, constraint handling, and uncertainty-aware control to prevent overconfident actuation under drift or partial observability.

*11) Full-Lifecycle Digital Twins and Continuous Digital Threads:* Moving beyond point solutions, next-generation Twins should establish an end-to-end Digital Thread spanning design, fabrication, assembly, field operation, and end-of-life. Lifecycle-aware integration of multimodal evidence enables holistic reliability governance: in-field degradation signatures can back-propagate to design rules and process windows, while remaining useful life (RUL) estimation can support maintenance planning, warranty strategy, and circular-economy decision-making.

*12) Security and Intellectual Property Protection:* Because Digital Twins consolidate sensitive design intent, material parameters, and process recipes, they constitute high-value targets for IP theft and integrity compromise. Secure, privacy-preserving collaboration frameworks are therefore essential for adoption across global supply chains. Federated learning, homomorphic encryption, secure enclaves, and differential privacy offer complementary mechanisms for enabling multi-stakeholder learning and data sharing without disclosing proprietary trade secrets or undermining model trustworthiness.

### G. Research Roadmap

A phased roadmap for advancing Digital Twins in 3D IC packaging is outlined across short-, medium-, and long-term horizons (summarized in Fig. 12). The intent is to sequence near-term enablers (data and latency), mid-term integration (standards and scalable deployment), and long-term autonomy (lifecycle synchronization and secure, closed-loop optimization).

**Short-term priorities (0–2 years): foundation and enablement.** Near-term progress should prioritize two immediate bottlenecks: data scarcity and computational latency. Key objectives include broader adoption of physics-informed learning architectures [41] to reduce labeled-data dependence and improve physical admissibility, alongside practical synthetic-data augmentation pipelines leveraging GANs, VAEs, diffusion models, and physics-based simulators to enrich rare-event coverage. In parallel, deployment-oriented efforts should emphasize edge-compatible implementations [42] and self-calibrating frameworks that fuse sparse in-situ sensing with high-fidelity simulation through principled data assimilation

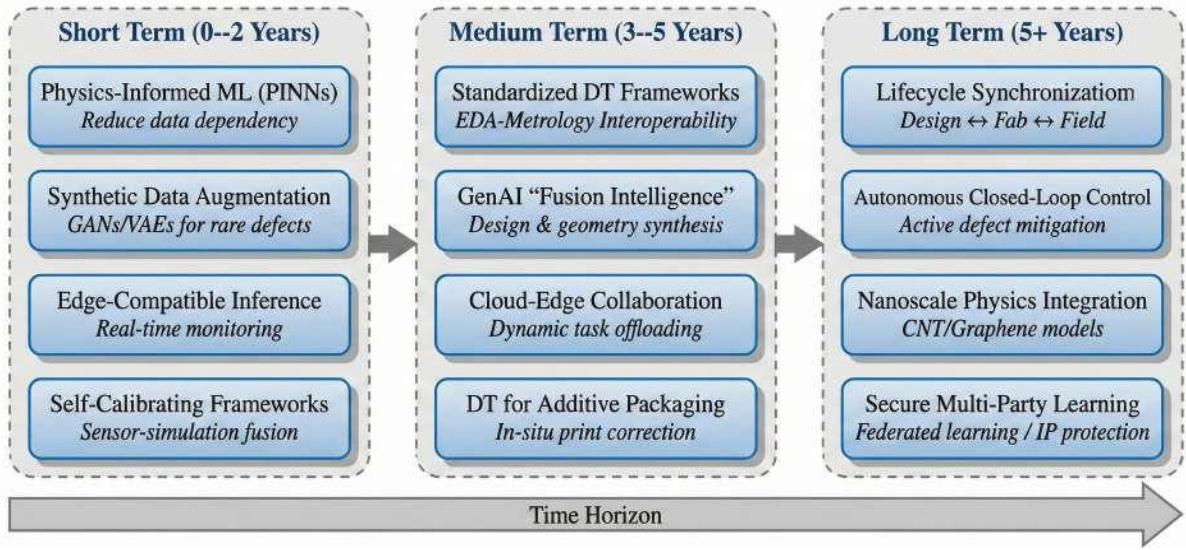

Fig. 12: Strategic research roadmap for Digital Twins in 3D IC packaging. The progression moves from foundation enabling technologies (Short Term) to system-level integration (Medium Term) and finally to fully autonomous, lifecycle-aware operation (Long Term).

[**?**], thereby establishing operational baselines for low-latency monitoring and inference.

**Medium-term priorities (3–5 years): integration and intelligence.** The medium-term focus shifts from standalone capability to interoperability and scalable system integration. Critical milestones include modular, standards-aligned DT frameworks that interface seamlessly with EDA environments, multiphysics solvers, and manufacturing telemetry, reducing reliance on brittle point-to-point pipelines. Architecturally, cloud–edge collaborative deployments with dynamic task offloading [40] should mature to support continuous model governance while meeting takt-time constraints. Methodologically, research should advance generative-AI-augmented DTs particularly "fusion intelligence" approaches [39] that combine generative modeling with physics-based guardrails and extend DT methodologies to additive manufacturing-enabled packaging, enabling predictive correction of warpage and layer-wise process stabilization in fine-pitch interconnect formation [43], [44].

**Long-term priorities (5+ years): autonomy and lifecycle synchronization.** The long-term objective is fully autonomous, lifecycle-aware Digital Twins capable of maintaining a continuous Digital Thread across design, fabrication, assembly, field operation, and end-of-life. Achieving this requires robust cross-stage synchronization, systematic uncertainty-aware decision support, and secure collaboration mechanisms that preserve IP while enabling multi-stakeholder learning. Technically, future Twins must incorporate emerging nanoscale interconnect and device physics [22], [45] and leverage distributed, low-latency ML infrastructures [46], [5] to support closed-loop control policies that proactively mitigate defect formation, manage degradation, and optimize yield and reliability at scale.

## IX. Conclusion

This review has systematically analyzed the emerging role of Digital Twins (DTs) within the 3D IC packaging ecosystem, categorizing the landscape into three distinct paradigms: machine learning (ML)-based, physics-based, and hybrid architectures. Our comparative analysis reveals a fundamental trade-off: while ML-based DTs excel in rapid, scalable defect screening, they often lack the mechanistic explainability required for safety-critical decisions. Conversely, physics-based models deliver high-fidelity insights into stress evolution and thermal management but are currently constrained by the computational cost of first-principles solvers. Hybrid DTs emerge as the most promising frontier, offering a pathway to reconcile data-driven efficiency with physics-based accuracy.

Despite the demonstrated potential of these technologies, widespread industrial adoption is currently impeded by significant barriers. Key challenges include the absence of standardized interoperability frameworks, the scarcity of high-quality failure data, and the difficulty of synchronizing models across the full product lifecycle—from design to end-of-life. Furthermore, the "real-time gap" remains a critical bottleneck, as current infrastructure struggles to support the low-latency feedback loops necessary for in-line process control.

To the best of our knowledge, this study represents the first comprehensive review dedicated specifically to Digital Twin applications in 3D IC packaging. By identifying the limitations of current paradigms and mapping out a phased research roadmap—ranging from synthetic data augmentation to self-calibrating, lifecycle-aware systems—this work aims to guide future research efforts. Ultimately, overcoming these barriers is essential for transitioning DTs from passive monitoring tools into active, intelligent agents capable of sustaining the reliability and performance of next-generation heterogeneous integration.

## REFERENCES

[1] "TSMC soic: Ultra-high-density 3d stacking technology." https://3dfabric.tsmc.com/english/dedicatedFoundry/technology/SoIC.htm, 2025.

[2] Advanced Micro Devices, Inc., "AMD 3D V-Cache™ Technology." Accessed: 2026-01-28.

[3] Samsung Electronics, "Samsung announces availability of its silicon-proven 3d ic technology for high-performance applications," Aug. 2020. Samsung Global Newsroom (Press Release), published Aug. 13, 2020. Accessed Jan. 28, 2026.

[4] IDTechEx, "Advanced semiconductor packaging 2025–2035: Forecasts, technologies, applications." https://www.idtechex.com/en/research-report/advanced-semiconductor-packaging/1042, 2024.

[5] A. Lancaster and M. Keswani, "Integrated circuit packaging review with an emphasis on 3d packaging," Integration, vol. 60, pp. 204–212, 2018.

[6] F. Che, X. Zhang, and J.-K. Lin, "Reliability study of 3d ic packaging based on through-silicon interposer (tsi) and silicon-less interconnection technology (slit) using finite element analysis," Microelectronics Reliability, vol. 61, 01 2016.

[7] J. Lau, "Overview and outlook of three-dimensional integrated circuit packaging, three-dimensional si integration, and three-dimensional integrated circuit integration," Journal of Electronic Packaging, vol. 136, 12 2014.

[8] EE Times Anon., "The multiphysics challenges of 3d ic designs." https://www.eetimes.com/the-multiphysics-challenges-of-3d-ic-designs/, 2024.

[9] A. Inamdar, P. Gromala, C. Bailey, L. Nguyen, B. Chan, J. Ryu, F. Rezaie, A. Detofsky, W. D. Van Driel, and G. Zhang, "Digital twins for ic packages and electronics-enabled systems," in 2024 IEEE Smart World Congress (SWC), pp. 2125–2132, 2024.

[10] N. Varshney, S. Ghosh, P. Craig, H. R. Kottur, H. Dalir, and N. Asadizanjani, "Challenges and opportunities in non-destructive characterization of stacked ic packaging: insights from sam and 3d x-ray analysis," p. 76, 10 2024.

[11] G. Haley, "Closing the test and metrology gap in 3d-ic packages." https://semiengineering.com/closing-the-test-and-metrology-gap-in-3d-ic-packages/, 2023.

[12] G. Shao, S. Jain, C. Laroque, L. H. Lee, P. Lendermann, and O. Rose, "Digital twin for smart manufacturing: the simulation aspect," in 2019 Winter Simulation Conference (WSC), pp. 2085–2098, IEEE, 2019.

[13] M. Grieves, "Origins of the digital twin concept," 08 2016.

[14] M. Shafto, M. Conroy, R. Doyle, E. Glaessgen, C. Kemp, J. LeMoigne, and L. Wang, "Draft modeling, simulation, information technology & processing roadmap," Technology area, vol. 11, pp. 1–32, 2010.

[15] A. Castellani, S. Schmitt, and S. Squartini, "Real-world anomaly detection by using digital twin systems and weakly supervised learning," IEEE Transactions on Industrial Informatics, vol. 17, pp. 4733–4742, July 2021.

[16] M. G. Kapteyn and K. E. Willcox, "Design of digital twin sensing strategies via predictive modeling and interpretable machine learning," Journal of Mechanical Design, vol. 144, p. 091710, 08 2022.

[17] S. Chen, P. V. Lopes, S. Marti, M. Rajashekarappa, S. Bandaru, C. Windmark, J. Bokrantz, and A. Skoogh, "Enhancing digital twins with deep reinforcement learning: A use case in maintenance prioritization," 2024 Winter Simulation Conference (WSC), pp. 1611–1622, 2024.

[18] M. Raissi, P. Perdikaris, and G. Karniadakis, "Physics-informed neural networks: A deep learning framework for solving forward and inverse problems involving nonlinear partial differential equations," Journal of Computational Physics, vol. 378, pp. 686–707, 2019.

[19] A. Quarteroni, P. Gervasio, and F. Regazzoni, "Combining physics-based and data-driven models: advancing the frontiers of research with scientific machine learning," Mathematical Models and Methods in Applied Sciences, vol. 35, p. 905–1071, Mar. 2025.

[20] Y. Zhou and S. J. Semnani, "A machine learning based multi-scale finite element framework for nonlinear composite materials," Engineering with Computers, Apr. 2025.

[21] N. Varshney, S. Ghosh, P. Craig, H. R. Kottur, H. Dalir, and N. Asadizanjani, "Challenges and opportunities in non-destructive characterization of stacked ic packaging: insights from sam and 3d x-ray analysis," Developments in X-Ray Tomography XV, vol. 13152, pp. 92–99, 2024.

[22] Y. M. Banadaki, A. Srivastava, and S. Sharifi, "Graphene nanoribbon field effect transistor for nanometer-size on-chip temperature sensor," in Nanosensors, Biosensors, and Info-Tech Sensors and Systems 2016, vol. 9802, pp. 12–20, SPIE, 2016.

[23] F. Altmann and M. Petzold, "Innovative failure analysis techniques for 3-d packaging developments," IEEE Design & Test, vol. 33, no. 3, pp. 46–55, 2016.

[24] J. M. Gu and S. J. Hong, "Virtual metrology for tsv etch depth measurement using optical emission spectroscopy," in 2015 IEEE Electrical Design of Advanced Packaging and Systems Symposium (EDAPS), pp. 27–30, IEEE, 2015.

[25] F. Kashfi and J. Draper, "Thermal sensor distribution method for 3d integrated circuits using efficient thermal map modeling," in 18th International Workshop on THERMal INvestigation of ICs and Systems, pp. 1–6, IEEE, 2012.

[26] J. C. Suhling, R. C. Jaeger, et al., "Silicon piezoresistive stress sensors and their application in electronic packaging," IEEE sensors journal, vol. 1, no. 1, pp. 14–30, 2001.

[27] C. Xi, A. A. Khan, J. True, N. Vashistha, N. Jessurun, and N. Asadizanjani, "Digital twin aided ic packaging structure analysis for high-quality sample preparation," in 2021 IEEE International Symposium on the Physical and Failure Analysis of Integrated Circuits (IPFA), pp. 1–6, 2021.

[28] A. Inamdar, M. van Soestbergen, A. Mavinkurve, W. van Driel, and G. Zhang, "Modelling thermomechanical degradation of moulded electronic packages using physics-based digital twin," Microelectronics Reliability, vol. 157, p. 115416, 2024.

[29] X. Wang, S. Cao, G. Lu, and D. Yang, "Viscoelastic simulation of stress and warpage for memory chip 3d-stacked package," Coatings, vol. 12, no. 12, 2022.

[30] Altair Engineering Inc., "Altair simlab: 3d digital twin for 3d ic and advanced ic packaging." https://altair.com/simlab/, 2025. Accessed: 2025-08-12.

[31] N. Dimitriou, L. Leontaris, T. Vafeiadis, D. Ioannidis, T. Wotherspoon, G. Tinker, and D. Tzovaras, "A deep learning framework for simulation and defect prediction applied in microelectronics," Simulation Modelling Practice and Theory, vol. 100, p. 102063, 2020.

[32] R. S. Pahwa, S. W. Ho, R. Qin, R. Chang, O. Z. Min, W. Jie, V. S. Rao, T. L. Nwe, Y. Yang, J. T. Neumann, R. Pichumani, and T. Gregorich, "Machine-learning based methodologies for 3d x-ray measurement, characterization and optimization for buried structures in advanced ic packages," in 2020 International Wafer Level Packaging Conference (IWLPC), p. 01–07, IEEE, Oct. 2020.

[33] Z. Liu, Y. Li, J. Hu, X. Yu, S. Shiau, X. Ai, Z. Zeng, and Z. Zhang, "Deepoheat: Operator learning-based ultra-fast thermal simulation in 3d-ic design," in 2023 60th ACM/IEEE Design Automation Conference (DAC), pp. 1–6, 2023.

[34] J. Luo, Y. Liu, K. Li, Z. Pan, C. Ma, and J. Lu, "Data-driven digital twin for board-level packaging interconnects under multi-physics loading," 03 2023.

[35] T. Wang, W. Zhong, X. Chen, Q. Ma, Y. Gu, W. Dong, and Z. Pan, "A self-calibrating digital twin approach by integrating cfd and sensor data for coal-fired boiler water wall temperature," Journal of Thermal Science, vol. 34, no. 3, pp. 738–755, 2025.

[36] F.-T. Cheng, H.-C. Huang, and C.-A. Kao, "Dual-phase virtual metrology scheme," IEEE Transactions on Semiconductor Manufacturing, vol. 20, no. 4, pp. 566–571, 2007.

[37] N. Asadi-Zanjani, T. John, and C. Xi, "Digital twin modeling of ic packaging structure," June 15 2023. US Patent App. 18/054,623.

[38] A. Inamdar, W. D. van Driel, and G. Zhang, "Digital twin technology—a review and its application model for prognostics and health management of microelectronics," *Electronics*, vol. 13, no. 16, 2024.

[39] R. Wang, M. Li, Z. Cao, J. Jia, K. Guan, and Y. Wen, "Fusion intelligence for digital twinning ai data centers: A synergistic genai-phyai approach," 2025.

[40] Z. Zhang, X. Zhang, G. Zhu, Y. Wang, and P. Hui, "Efficient task offloading algorithm for digital twin in edge/cloud computing environment," 2023.

[41] M. S. Zia, C. Houpert, A. Anjum, L. Liu, A. Conway, and A. Peña-Rios, "Physics encoded blocks in residual neural network architectures for digital twin models," *Machine Learning*, vol. 114, no. 8, p. 180, 2025.

[42] S. K. Das, M. H. Uddin, and S. Baidya, "Edge-assisted collaborative digital twin for safety-critical robotics in industrial iot," *arXiv preprint arXiv:2209.12854*, 2022.

[43] D. Gamdha, K. Saurabh, B. Ganapathysubramanian, and A. Krishnamurthy, "High-resolution thermal simulation framework for extrusion-based additive manufacturing of complex geometries," 2025.

[44] V. Karkaria, A. Goeckner, R. Zha, J. Chen, J. Zhang, Q. Zhu, J. Cao, R. X. Gao, and W. Chen, "Towards a digital twin framework in additive manufacturing: Machine learning and bayesian optimization for time series process optimization," 2024.

[45] A. Srivastava, X. H. Liu, and Y. M. Banadaki, *Overview of Carbon Nanotube Interconnects*, pp. 37–80. Cham: Springer International Publishing, 2017.

[46] Y. Banadaki and S. Sharifi, "Cyber-enabled distributed machine learning for smart manufacturing systems," in *Smart Structures and NDE for Energy Systems and Industry 4.0*, vol. 10973, pp. 193–198, SPIE, 2019.